\documentclass[elsarticle]{article}
\usepackage[title]{appendix}
\usepackage[affil-it]{authblk}
\usepackage[intlimits]{amsmath}
\usepackage{amsfonts}
\usepackage{wrapfig}

\usepackage{graphicx}
\usepackage{caption}

\graphicspath{ {C:\Users\chris\Desktop\LETTER} }
\usepackage{subfigure}
\usepackage[usenames]{color}
\usepackage{epstopdf,epsfig,todonotes}
\usepackage{xcolor}
\usepackage{indentfirst}
\usepackage{enumerate}
\usepackage{tikz}
\usepackage{mathtools}
\usepackage{graphicx}
\usepackage{amssymb, mathtools, mathrsfs}
\usepackage{comment}
\usepackage{hyperref}

\newcommand\be            {\begin{equation}}
\newcommand\ee            {\end{equation}}
\def\d{{\rm d}}

\def\bea{\begin{eqnarray}}
\def\eea{\end{eqnarray}}
\newcommand{\C}{\mathbb{C}}

\newcommand\EN           {\end{equation}}
\newcommand\bes           {\begin{subequations}}
\newcommand\esu           {\end{subequations}}

\newcommand{\ol}{\overline}

\newcommand{\R}{\mathbb{R}}

\def\3pt#1#2#3{{\langle{#1}\vert{#2}\vert{#3}\rangle}}
\newcommand\dsl[1]{{{\displaystyle{#1}}}}
\DeclareMathOperator{\im}{\dot{\imath}} 

\usepackage{amsmath,amssymb}

\begin{document}

\title{Shielding of  breathers for  the focusing nonlinear Schr\"odinger equation}

\author{Gregorio Falqui,${}^{a,b}$ \, Tamara Grava,${}^{c,d,e} \,  $Christian Puntini${}^{f}$ }
\affil{${}^a$ Dipartimento di Matematica e Applicazioni, Universit\`a di Milano-Bicocca, via R. Cozzi 55, 20125 Milano, Italy\\
\medskip  ${}^b$ INFN, Sezione di Milano,\, 
${}^c$ INFN, Sezione di Trieste,\\
\medskip
${}^d$ SISSA, via Bonomea 265, 34136, Trieste, Italy\\
\medskip 
${}^e$ School of Mathematics, University of Bristol, Fry Building, Bristol, BS8 1UG, UK\\ \medskip 

 ${}^f$ Faculty of Mathematics, University of Vienna, Oskar-Morgenstern-Platz 1, 1090 Vienna,	Austria.}
\maketitle

\begin{abstract} 

\noindent
We study a deterministic gas of  breathers for the Focusing Nonlinear Schr\"odinger 
equation. The gas of breathers is obtained from  a $N$-breather solution in  the limit  $N\to \infty$. 
The  limit is performed at the level of scattering data by letting  the $N$-breather spectrum to  fill uniformly a suitable compact domain
of the complex plane  in the limit $N\to\infty$. The  corresponding norming constants  are  interpolated by a smooth function   and scaled as $1/N$.
For particular choices of the domain and the interpolating function, 
 the   gas of breathers  behaves as  finite  breathers solution.   This extends the  \textit{shielding effect} discovered in Bertola, Grava, Orsatti (2023)  \cite{Bertola_G_O_2022} for a soliton gas also to  a breather gas.
  \end{abstract}

\noindent
\newtheorem{theorem}{Theorem}
\newtheorem{lemma}{Lemma}
\newtheorem{corollary}{Corollary}
\newtheorem{prop}{Proposition}
\newenvironment{proof}{{\bfseries Proof.}}{\qedsymbol}
\newcommand{\qedsymbol}{$\blacksquare$}

\newtheorem{remark}{Remark}


			
	\section{Introduction}
The hallmark of integrability of a non-linear dispersive wave equation is the existence of pseudo-particle solutions,  that is, coherent nonlinear modes that interact elastically.    Recently several investigations  involving very large number of solitons  have been carried out, both on the mathematical and the physical side (\cite{El_T_2020},
 \cite{GGJM}, \cite{GGJMM},  \cite{Bilman_M_2017},  \cite{GASR21}, \cite{Gelash19}, \cite{Roberti2021}). Coherent and random superpositions of solitons were considered. 

Random nonlinear superpositions of solitons are more closely related to the notion of a soliton gas in an infinite statistical ensemble of interacting solitons that was first introduced by Zakharov by considering the Korteweg de Vries (KdV) equation in \cite{Zakharov1971}. More recently {this concept has been} extended to the Focusing Nonlinear Schr\"odinger  (FNLS) equation by El and Tovbis in \cite{El_T_2020}{, and has been studied in the context of the modified Korteweg-de Vries (mKdV) equation with nonzero boundary conditions by Zhang and Ling in \cite{ZhangLing}}. {Morevoer}, in \cite{Bertola_G_O_2022} the problem of both a deterministic and a random distribution of $N$ soliton solutions of FNLS was discussed and in \cite{GGMN} a central limit theory for a soliton gas has been derived.
 In \cite{Bertola_G_O_2022} it was proved that a particular configuration of a  deterministic soliton gas surprisingly yields a  one-soliton solution. Such a phenomenon was termed {\em soliton shielding} and further studied in \cite{BGO2024}.

In this paper we study the corresponding problem considering the case of a {\em gas of breathers}. Generally speaking, a breather is a nonlinear wave in which energy concentrates in a localized and oscillatory way.   The first breather type solution of FNLS was found in \cite{Kuznetsov_1977} and  \cite{Ma_1979} and  called {\em Kuznetsov-Ma breather}. It is localized in the
spatial variable $x$ and periodic in the temporal variable $t$.
 In particular, in \cite{Ma_1979} the initial value problem for FNLS with  initial data  a perturbed plane wave solution with non-vanishing boundary conditions at  $x\rightarrow \pm \infty$ was shown to yield  asymptotic states consisting  of periodically ``pulsating" solitary waves  (plus a residual of small amplitude dispersive waves).
 
Different kind of such pulsating localized solution to FNLS were subsequently discovered. In~\cite{Peregrine_1983} a rational solution of this kind, called {\em Peregrine breather}  was found. It  is localized in both spatial and temporal variables $(x, t)$. In~\cite{Akhmediev1987}  an $x$-periodic and $t$-localized new solution, called {\em Akhmediev breather} was discovered. In 1988 Tajiri and Watanabe \cite{Tajiri_W_1998} found  a travelling breather, now called {\em Tajiri-Watanabe breather}, whose speed depends on the spectrum of the associated linear problem. Such solutions have been   shown to be a model of modulational instability by Zakharov and Gelash in 2013 (\cite{Zakharov2013} and \cite{35}).

The Kuznetsov-Ma, Peregrine and Akhmediev breathers find applications as prototypical models of rogue waves in nonlinear optics and  the theory of gravity water waves  \cite{19}, \cite{21}, \cite{DystheTrulsen1999}, \cite{CostaOsborne2014}, \cite{Gelash2018}.   Furthermore the Peregrine breather emerges in a universal way in the semiclassical limit of the focusing nonlinear Schr\"odinger equation \cite{BT}.
Higher order breathers solution of FNLS  obtained using Darboux transformations from the rational Peregrine breather in \cite{Akhetal1}, \cite{Akhetal2} were further studied in \cite{Bilman_M_2017}  and observed  in a water-wave tank to be approximation of ``super rogue waves" with an amplitude of up to $5$ times the background in \cite{Chabchoub2012}.
 This analysis was pushed forward by Bilman, Buckingham,  Ling, Miller  and Wang    \cite{BM2024} \cite{BLM2020}, \cite{BB2019},\cite{BM2021},  \cite{BBW}    where breathers and solitons of infinite order have been considered, 
 and a universal profile emerges after an appropriate scaling.

The starting  point of our study is the fact that 
 $N$-breather solutions to FNLS  can be obtained from the Zakharov-Shabat  linear spectral problem, reformulated as a Riemann-Hilbert problem  for a $2\times 2$ matrix $M(z;x,t)$, $z$ being the spectral parameter, with suitable scattering data~\cite{Bilman_M_2017}, \cite{Biondini_K_2014}. In particular, for the  Kuznetsov-Ma and  the Tajiri-Watanabe  $N$-breather, one needs $4N$  poles where $N$ poles are free parameters located in the upper half complex plane minus the unit half disk, while the remaining $3N$ poles are obtained by suitable symmetries.  The spectral data are completed by assigning to each of the first $N$ poles a norming constant,  while the remaining norming constants are obtained by symmetry requirements. Our main aim is to study the limit of the $N$-breather solution when the number $N$ goes  to infinity and the location of the poles accumulates  uniformly  on  a  bounded domain $\mathcal{D}_1$ of the complex plane (and its ``symmetric" counterparts as required by the very notion of breather solution).   The corresponding norming constants are interpolated by a smooth function and rescaled by $N$. In this way we arrive at the inverse problem for a  gas of  breathers   and  we show the existence of its solution.
For certain choices of the domain $\mathcal{D}_1$  and the interpolation function,   the behaviour of the  gas of  breathers  corresponds to a    $n$-breather solution, $n$ being determined by the geometry of the above-mentioned domain  of pole accumulation $\mathcal{D}_1$.

The layout of the present paper is the following. 
\begin{itemize}
\item In the second section  we briefly review the characteristic properties of the inverse spectral problem for breather solutions, as well as the associated Riemann-Hilbert problem, and frame the Kuznetsov-Ma and the Tajiri-Watanabe 1-breather solution within this scheme. 
\item  In section three we transform the Riemann-Hilbert problem for $N$ poles into a Riemann-Hilbert problem with a jump matrix defined on closed contours encircling the accumulation domains of the poles and describe its solution.
\item  We finally pass to  the $N\to\infty $ limit, showing the existence of the solution of the limiting inverse problem. We  illustrate the shielding effect for particular choices of the domain  $\mathcal{D}_1$ and we  explicitly illustrate our procedure and results for the cases  when the limiting solution reduces to a  $n=1$ and $n=2$ shielding breather, corresponding to the choice of   the domain  $\mathcal{D}_1$ (respectively)  a disk and an eight-shaped figure.
\end{itemize}

	\section{A brief review of the IST for breathers}		
			\noindent
			In this section we report and summarize the main steps of  the inverse scattering transform for the $N$-breather solution of the FNLS equation 
			\begin{equation}\label{FNLS}
				\dot{\imath}\psi_t + \psi_{xx}+2|\psi|^2\psi=0\, , 
			\end{equation}
with non-zero boundary conditons at $x=\pm \infty$ originally introduced {in \cite{Kuznetsov_1977}} and further developed in \cite{Biondini_K_2014}, \cite{Biondini}.\\
			In order to remove the possible time-dependence of the boundary conditions, we can rewrite equation  \eqref{FNLS} as 
			\begin{equation}\label{biondini 1.1}
				\dot{\imath}\psi_t + \psi_{xx} + 2 \left(|\psi|^2-\psi_0^2\right)\psi=0\, , 
			\end{equation}
		the nonzero boundary conditions as $x\rightarrow \pm \infty$ being
			\begin{equation}\label{biondini 1.2}
				\lim_{x\rightarrow \pm \infty} \psi(x,t)=\psi_{\pm},\quad |\psi_{\pm}|=:\psi_0\not=0.
			\end{equation}
			The additional term $-2\psi_0^2\psi$ in \eqref{biondini 1.1} that has been  added in order to make the boundary conditions \eqref{biondini 1.2} independent of time  can be removed by the rescaling $\widetilde{\psi}(x,t)=    e^{2\dot{\imath}\psi_0^2 t}\psi(x,t)$.\\
			As it well known, the FNLS equation  \eqref{FNLS} admits the scaling invariance  
				\begin{equation}
					\psi(x,t)\to \sigma\psi( \sigma x, \sigma^2 t), \quad \forall\  \sigma >0,
				\end{equation} 
				{hence}, we may assume that $\psi_-=\psi_0=1$ and $\psi_+$  arbitrary constant with $|\psi_+|=1$.
				The  nonlinear Schr\"odinger equation is the compatibility condition of the 
				 Zakharov-Shabat   linear  spectral problem   \cite{Shabat1972}
				\begin{equation}
\label{eq:ZSsystem}
\begin{split}
& \partial_{x}f = \mathcal{L}f, \qquad \mathcal{L} =ik{ \sigma}_3+ \Psi,\quad   \Psi= \begin{pmatrix}0 & \psi \\ -\overline{\psi} & 0\end{pmatrix}, \\ 
&i \partial_{t}f = \mathcal{B}f, \qquad \mathcal{B} = -2ik^2{ \sigma}_3 + i \sigma_3(  \Psi_x-\Psi^2- 1)-2k \Psi \ ,
\end{split}
\end{equation}
where $ \sigma_3=\begin{pmatrix}1&0\\0&-1\end{pmatrix}$ and $ \overline{\psi}$ stands for complex conjugate. The direct and inverse spectral problems with non zero boundaries conditons have been {developed by Biondini and Kova\v{c}i\v{c} in \cite{Biondini_K_2014}, extending the seminal work of Kuznetsov in 1977 \cite{Kuznetsov_1977}.  Here,} we summarize  the inverse spectral problem for reflectionless potentials in the the complex $z$-plane where  $z=k+\sqrt{k^2+1}$.												\noindent
\begin{figure}[h!]						\centering
						\includegraphics[width=8cm]{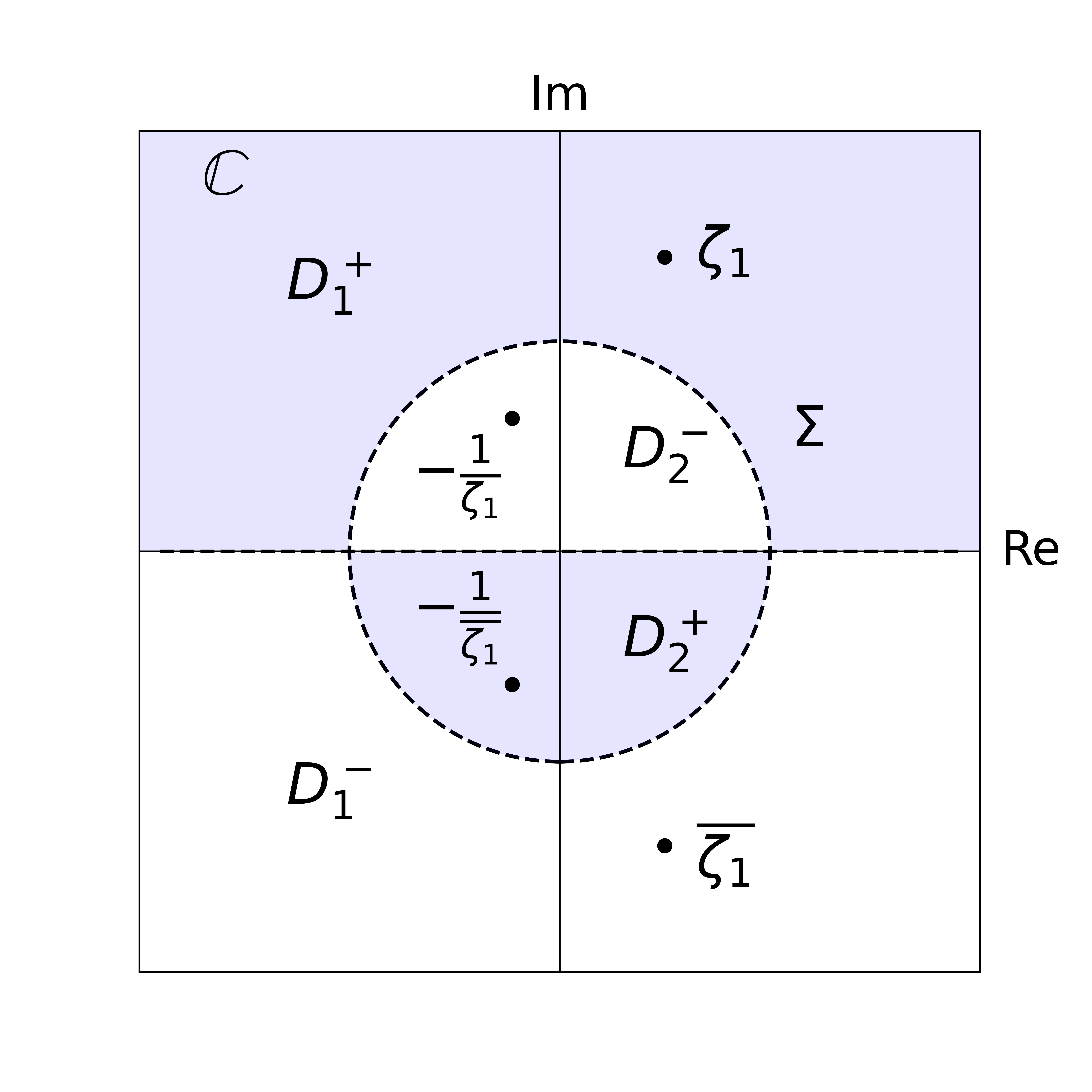}
						\caption{The complex plane divided as in \eqref{piano complesso diviso} and 4 poles satisfying \eqref{symmetrie poli}. Fixing $\zeta_1$ in $D_1^+$ determines the other three ``fellow" points of the spectrum, $\zeta_2=-1/{\bar{\zeta_1}},\zeta_3= \bar{\zeta_1},\zeta_4=-{1}/{\zeta_1}$.}
						\label{piano complesso diviso jpg}
					\end{figure}

		\noindent
		The spectral $z$-complex plane  is partitioned  into  the following domains:
			\begin{equation}\label{piano complesso diviso}
				\begin{split}
				& D^+=\left\{ z \in \mathbb{C}\ :\ \left(|z|^2-1\right)\Im(z) >0 \right\},\\
				& D^-=\left\{ z \in \mathbb{C}\ :\ \left(|z|^2-1\right)\Im(z) <0 \right\},
			\end{split}	
			\end{equation}
			
			with boundary
				$\Sigma=\mathbb{R}\cup \mathsf{S}^1=\mathbb{R}\cup\left\{ z \in \mathbb{C}\ :\ |z|=1\right\}$.
			We observe  that $D^+$ is the union of the exterior to the unit disk in the upper half plane $D^+_1$ and the interior of the unit disk in the lower half plane $D^+_2$.  The domain  $D^-$, obtained   from  $D^+$  by  reflection along the real axis,  is equally partitioned into an unbounded domain $D^-_1$ in the lower half plane and a bounded domain $D^-_2$ in the upper half plane (see Figure \ref{piano complesso diviso jpg}).\\
The $N$-breather solution {is obtained from}  the following  scattering  data:\\ the discrete spectrum   $\mathcal{Z}=\{\zeta_1,...,\zeta_{2N},\overline{\zeta_1},...,\overline{\zeta_{2N}}\}$  of the Zakharov-Shabat linear spectral problem \cite{Biondini_K_2014}  with 
					{\begin{equation}\label{symmetrie poli}
						\begin{aligned}
							\zeta_j &\in D^+_1, & \zeta_{N+j} &= -\frac{1}{\overline{\zeta_j}} \in D^+_2, \\
							\overline{\zeta_j} &\in D^-_1, & \overline{\zeta_{N+j}} &= -\frac{1}{\zeta_j} \in D^-_2, 
						\end{aligned} \quad \text{with}\ j = 1, \ldots, N.
					\end{equation}}
		The points of the discrete spectrum satisfy, with respect to the boundary values $\psi_\pm $   in \eqref{biondini 1.2}, the ``theta" condition  :
			\begin{equation}\label{biondini 3.14 theta cond breather}
				\arg\left(\frac{\psi_-}{\psi_+}\right)=4\sum_{j=1}^N\arg(\zeta_j).
			\end{equation}
	{Condition \eqref{biondini 3.14 theta cond breather} is termed ``theta" as it relates the discrete spectrum, the elements of the scattering matrix and the asymptotic phase difference of the corresponding FNLS solution, originally denoted  by $\theta$ in \cite{FaddeevTakhtajan} .
		  As we are assuming  $\psi_-=1$, \eqref{biondini 3.14 theta cond breather} becomes, in our case, 			\begin{equation}\label{theta cond breather}
		-\arg\left({\psi_+}\right)=4\sum_{j=1}^N\arg(\zeta_j),\quad (\text{mod}\ 2\pi).
		\end{equation} }
			To complete the  scattering data, we need to add the norming constants. For each point of the spectrum 	{$ \{\zeta_j\}_{j=1}^N$ in $D^+_1$,}
		 we chose a nonzero complex number, obtaining the sequence  $\{C_j\}_{j=1}^N $.
			The remaining {$3N$ norming} constants are obtained by symmetry  as follows: for points in {$D^+_2\subset D^+$} one sets			
			\begin{equation}\label{biondini norming constants 1}
				 C_{N+j}=-\left(\frac{1}{\overline{z_j}}\right)^2\overline{C_j},\quad \text{for } j=1,...,N,
			\end{equation}
			while the $2N$ norming constants associated with the poles in $D^-$ are obtained from the previous ones as
			\begin{equation}\label{biondini norming constants 2}
				{C}_{2N+\alpha}=-\overline{C_\alpha},\quad \text{for}\ \alpha=1,...,2N.
			\end{equation}
			\bigskip
			The  {$N$-breather solution is} recovered    by solving the following  Riemann-Hilbert problem.\\	\medskip
			{{\textbf {Riemann-Hilbert problem A (Rational $N$-breather problem):}} \\		To find a $2\times2$ matrix $M(z;x,t)$ with the following properties:
			\begin{itemize}\item 		$M(z;x,t)$ is a meromorphic matrix 	in $\C$, with simple poles  in the set $\mathcal{Z}=\{\zeta_{\alpha}, \overline{\zeta_{\alpha}}\}_{\alpha=1}^{2N}$ and $z=0$.	
			\item 		$M(z;x,t)$  satisfies the following residue conditions			
			\begin{equation}\label{biondini 3.8}
				\begin{split}
				\mathop{\mathrm{Res}}_{z = \zeta_\alpha}M(z;x,t)=\lim_{z\rightarrow\zeta_\alpha}M(z;x,t)\left[\begin{array}{cc}
					0&0  \\
					C_\alpha e^{-2\dot{\imath}\theta\left(\zeta_\alpha ;x,t\right)} & 0
				\end{array}\right], \\
				\mathop{\mathrm{Res}}_{z =\overline{\zeta_\alpha}}M(z;x,t)=\lim_{z\rightarrow\overline{\zeta_\alpha}}M(z;x,t)\left[\begin{array}{cc}
					0&-\overline{C_{\alpha}}e^{2\dot{\imath}\overline{\theta\left(\zeta_\alpha;x,t\right)}}\\
					0& 0
				\end{array}\right],
			\end{split}
			\end{equation} 
	for $ \alpha=1,...,2N$,	with
			\begin{equation}
			\theta(z;x,t)=\frac{1}{2}\left(z+\frac{1}{z}\right)\left(x -\left(z-\frac{1}{z}\right)t\right)\,.
		\end{equation}
			\item $M(z; x, t)$  satisfies  the   following asymptotic conditions: 
			\begin{align}\label{asymptotic}
				& M(z;x,t)=\mathbb{I} + \mathcal{O}\left(\frac{1}{z}\right), \quad\text{as}\ z\rightarrow \infty,\\
				\label{asymptotic1}
				& M(z;x,t)=\frac{\dot{\imath}}{z}\sigma_1 + \mathcal{O}\left(1\right), \quad\text{as}\ z\rightarrow 0\,  ,
			\end{align}
			where $\sigma_1$ is the Pauli matrix  $\begin{bmatrix} 0 &1  \\
									1 & 0\end{bmatrix}$. 						
\end{itemize}
}
\noindent
For further use we note  that 
		\begin{equation}	\label{simmetria theta}
			\overline{\theta(z;x,t)}=\theta(\overline{z};x,t),\quad \text{and }\quad
\theta\left(-\frac{1}{z};x,t\right)=-\theta(z;x,t).
		\end{equation}
\noindent
				From the solution of the above Riemann-Hilbert problem A for  the $2\times 2$ matrix $M$, one recovers the  $N$-breather  solution of the 	
				focusing nonlinear Schr\"odinger equation by the relation
				\begin{equation}
				\label{NLS_sol}
				\psi(x,t)=-i\lim_{z\to\infty}(zM_{12}(z;x,t)),
				\end{equation}
				where $M_{12}$ is the $1,2$ entry  of  $M$.\\
The conditions \eqref{biondini 3.8} show that the matrix $M$ is meromorphic in the complex plane;  in particular, for $z\in D_+$, the second column of $M$ is analytic while the first column has a first order pole at $ \zeta_\alpha\in D_+$, $\alpha=1,\dots, 2N$.
On the other hand the second column of $M$ is analytic in $D_-$ and it has a  first order pole at $\overline{\zeta_\alpha}\in D_-$, $\alpha=1,\dots, 2N$. \\
Taking into account the asymptotic conditions \eqref{asymptotic},
 the $2\times 2$ matrix $M(z;x,t)$   takes the form 
 			\begin{equation}\label{Martix M}				
			M(z;x,t)=\mathbb{I} + \frac{\dot{\imath}}{z}\left[\begin{array}{cc}
									0 &1  \\
									1& 0
								\end{array}\right]\ +\sum_{\alpha=1}^{2N}\frac{\left[\begin{array}{cc}
						A_\alpha(x,t)  & 0 \\
						B_\alpha(x,t) & 0
					\end{array}\right]}{z-\zeta_\alpha}+\sum_{\alpha=1}^{2N}\frac{\left[\begin{array}{cc}
						0  & E_\alpha(x,t)  \\
						0 &  F_\alpha(x,t)
					\end{array}\right]}{z-\overline{\zeta_\alpha}},			
			\end{equation}
			where the coefficients $A_\alpha(x,t)$, $B_\alpha(x,t)$, $E_\alpha(x,t)$, $F_\alpha(x,t)$ are to be determined from the residue conditions \eqref{biondini 3.8} as follows.  Let us introduce the notation
		\[
			c_\alpha=C_\alpha e^{-2\dot{\imath}\theta\left(\zeta_\alpha ;x,t\right)},\quad \alpha=1.\dots, 2N\,.
			\]
			From the residue condition \eqref{biondini 3.8} of the $N$-breather solution   and the ansatz \eqref{Martix M}    we obtain
				\begin{equation}
				\label{sys_tot}
				\begin{split}
				&\begin{bmatrix}
				A_\beta\\
				B_\beta
				\end{bmatrix}=\begin{bmatrix}\frac{ic_\beta}{\zeta_\beta}\\c_\beta\end{bmatrix}+c_\beta\sum_{\alpha=1}^{2N}\dfrac{\begin{bmatrix}
				E_\alpha\\
				F_\alpha
				\end{bmatrix}}{\zeta_\beta-\overline{\zeta_\alpha}},\\
				&\begin{bmatrix}
				E_\beta\\
				F_\beta
				\end{bmatrix}=-\begin{bmatrix}c_\beta\\ \frac{ic_\beta}{\zeta_\beta}\end{bmatrix}-\overline{c_\beta}\sum_{\alpha=1}^{2N}\dfrac{\begin{bmatrix}
				A_\alpha\\
				B_\alpha
				\end{bmatrix}}{\overline{\zeta_\beta}-\zeta_\alpha},
				\end{split}		
			\end{equation}
			for $\beta=1,\dots, 2N.$
This is a system of $2N$ equations in $2N$ unknowns $A_\alpha, B_\alpha, E_\alpha$ and $F_\alpha$ (see  Appendix ~\ref{AppA}  for a comprehensive  solution).  
		
	From \eqref{NLS_sol} {and \eqref{sys_tot}}, the $N$-breather solution of the focusing nonlinear Schr\"odinger equation is finally recovered as
		\begin{equation}\label{Nbreathers}
		\psi(x,t)=-\im \lim_{z\rightarrow \infty} z M_{12}(z;x,t)=1-\im\sum_{\alpha=1}^{2N}E_\alpha(x,t).
	\end{equation}

	\subsection{Kuznetsov-Ma and Tajiri-Watanabe breathers}
	\noindent
	We report  the formul\ae\ of the Kuznetsov-Ma  \cite{Kuznetsov_1977}, \cite{Ma_1979} and the Tajiri-Watanabe   \cite{Tajiri_W_1998}, \cite{Agafontsev2024} breather solutions. 	The formul\ae\  presented here have been derived  from the expression $\psi(x,t)=1-\im (E_1(x,t)+E_2(x,t))$ where $E_1(x,t)$ and $E_2(x,t)$ solve  the linear system  \eqref{lin_2} in Appendix A. 
	The Kuznetsov-Ma breather is obtained from the data $(\zeta_1, C_1)$ where  the single pole   $\zeta_1$ is purely imaginary  
	\begin{equation}\label{z_1}
		\zeta_1=\dot{\imath}Z, \;\; \,  Z>1\, \;\;\mbox{and  } C_1=e^{\kappa+\dot{\imath}\phi},\quad \mbox{with  $ \kappa,\ \phi$ real}.
		\end{equation}
		The remaining poles  and norming constants  are obtained by symmetry according to \eqref{symmetrie poli},
		\eqref{biondini norming constants 1} and \eqref{biondini norming constants 2}  respectively.	
%
The Kuznetsov-Ma breather solution is {given by}
	\begin{equation}\label{biondini KM}
		\psi_{1KM}(x,t)=\frac{\cosh{(\chi(x))}-\frac{1}{2}c_+\left(1+{c_-^2}/{c_+^2}\right)\sin{(s(t))}+\dot{\imath}c_-\cos{(s(t))}}{\cosh{(\chi(x))}-A\sin{(s(t))}},
	\end{equation}
	with
		\begin{equation}\label{biondini 4.2 KM}
	\begin{split}
					{c_{\pm}}&=Z\pm\frac{1}{Z},\quad A=\frac{2}{{c_+}}<1,\\ 			
			{\chi}(x)&={c_-}x+{c_0}+\kappa,\quad{c_0}=\ln\left(\frac{c_+}{2Zc_-}\right), \\
			s(t)&={c_+}{c_-}t-\phi.
		\end{split}
	\end{equation}
\noindent
{We remark that the} Kuznetsov-Ma breather is a stationary breather, since it is localized in the
spatial variable $x$ and periodic in the temporal variable $t$, with period $T =\frac{2\pi}{c_+c_-}=\frac{2\pi}{Z^2-1/Z^2}$.
\begin{figure}[h!]
		\centering
		\includegraphics[width=.46\linewidth]{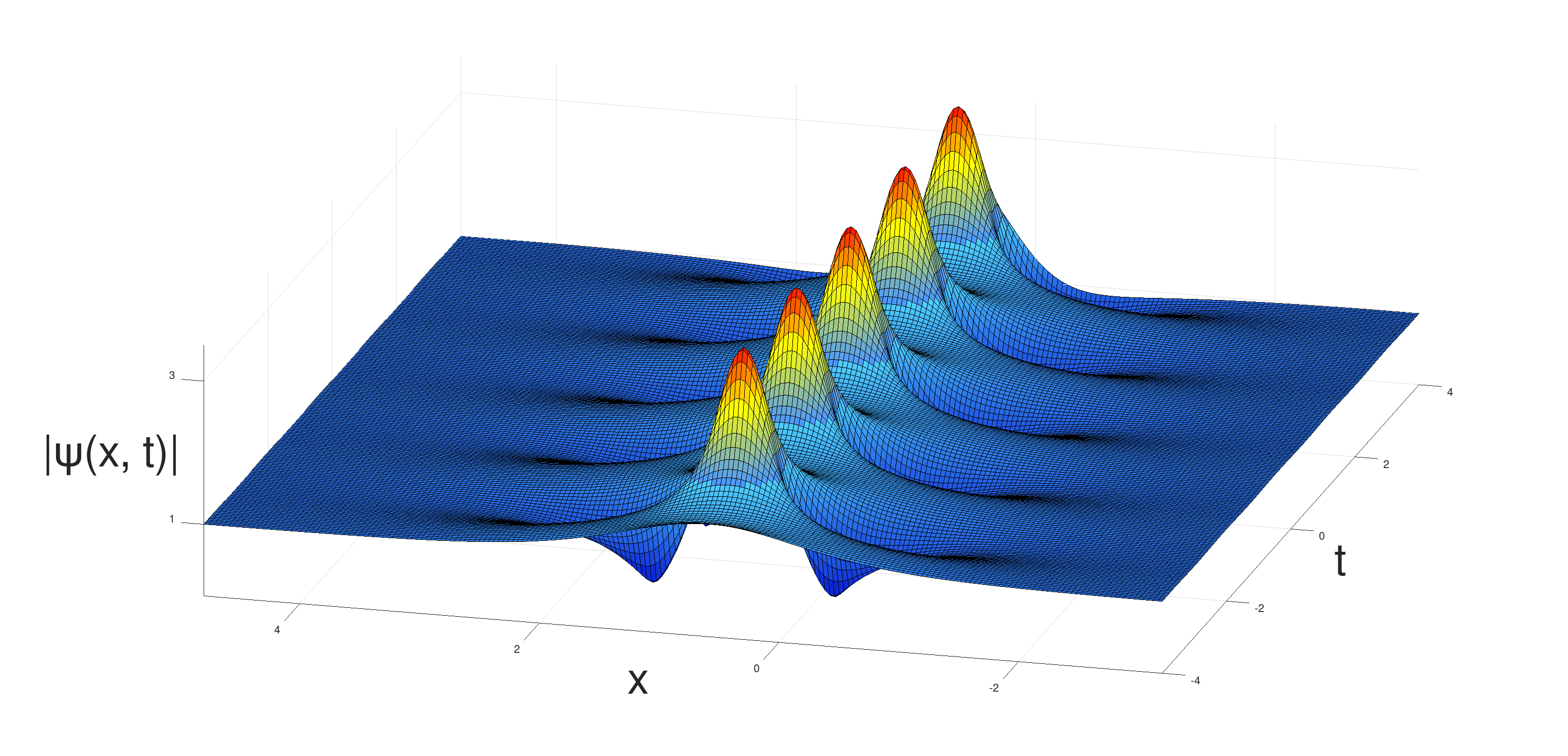}
		\centering
		\includegraphics[width=.46\linewidth]{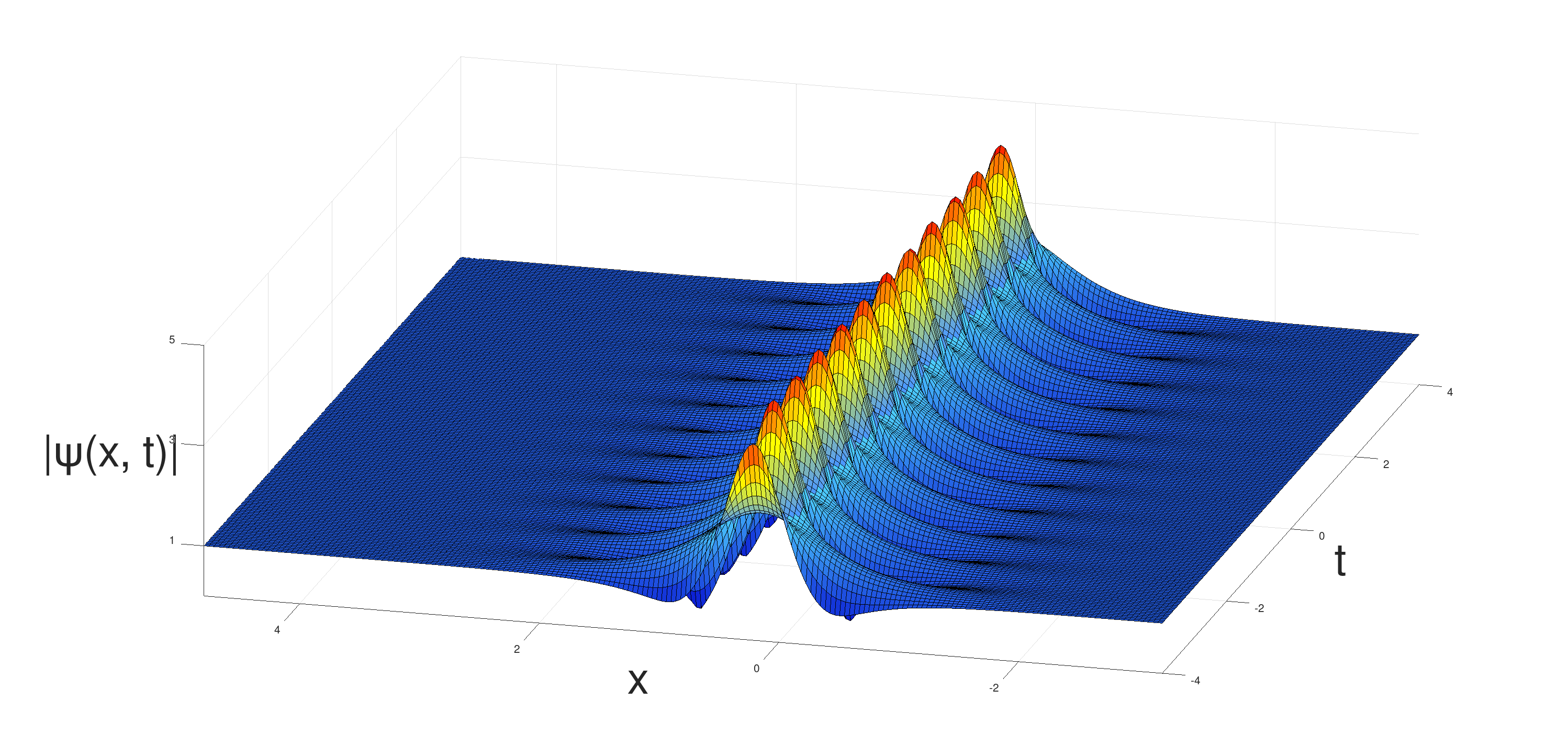}
	\caption{Two different Kuznetsov-Ma breathers  with $Z=2$, $\kappa=0$, $\phi=0$  (left)  and  $Z=3$, $\kappa=1$, $\phi=0$ (right).  In both cases, $\psi(x,t)\to 1$ as  $x\to\pm \infty.$}
\end{figure}
\newline
\noindent
The Tajiri-Watanabe breather is instead obtained by choosing
\begin{equation}
\begin{split}
	\zeta_1&=\im Z e^{\im \alpha}, \quad \text{with}\ Z>1\quad\text{and}\ -\frac{\pi}{2}<\alpha<\frac{\pi}{2},\\
	C_1&=e^{\kappa+\dot{\imath}\phi},\quad \mbox{with $\kappa$ and $\phi$ real. }
	\end{split}
\end{equation}
The corresponding formula for the {solution} is 
\begin{equation}\label{biondini tw 4.5}
		\psi_{1TW}(x,t)=\frac{\cosh (\chi -2\im \alpha) +\frac{B}{2}\left[d_+(Z^2\sin(s-2\alpha)-\sin(s))+\im d_-(Z^2 \cos(s-2\alpha) -\cos(s)\right]}{e^{2\im \alpha}(\cosh (\chi)+B\left[Z^2 \sin(s-2\alpha)-\sin(s)\right])} ,
\end{equation}
with
\begin{equation}
	\begin{gathered}
		 c_{\pm}=Z\pm\frac{1}{Z}, \quad d_{\pm }=Z^2\pm\frac{1}{Z^2},\\
		B=\frac{2\cos(\alpha)}{|1-Z^2 e^{-2\im \alpha}|\left(Z+\frac{1}{Z}\right)},\\		 
	\chi(x,t)=c_- x \cos(\alpha) + d_+t \sin(2\alpha) + c_0' + \kappa,\\
	 s(x,t)=c_+ x \sin(\alpha) - d_{-} t \cos(2\alpha) + \phi,\\
	 c_0'=-\ln\left(\frac{2\cos(\alpha)|1-Z^2 e^{-2\im \alpha}|}{Z+\frac{1}{Z}}\right)	\,.		
 \end{gathered}
\end{equation}
Let us observe that 
\[
\psi_{1TW}(x,t)\to 1\; \mbox{as $x\to-\infty$},\quad \psi_{1TW}(x,t)\to e^{-4\im\alpha}, \;\; \mbox{as $x\to+\infty$}.
\]

\begin{figure}[h!]
		\centering
		\includegraphics[width=.45\linewidth]{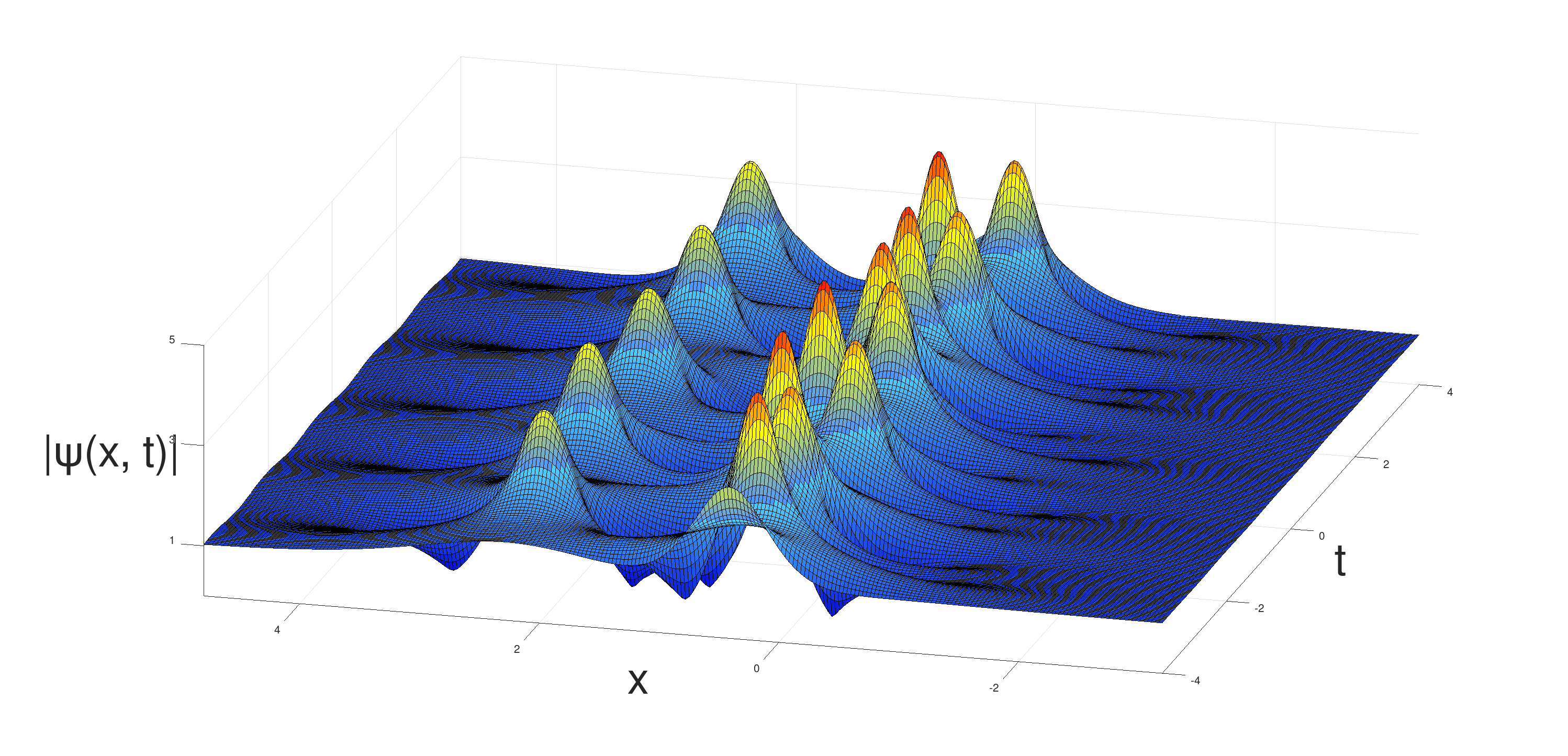} \includegraphics[width=0.45\linewidth]{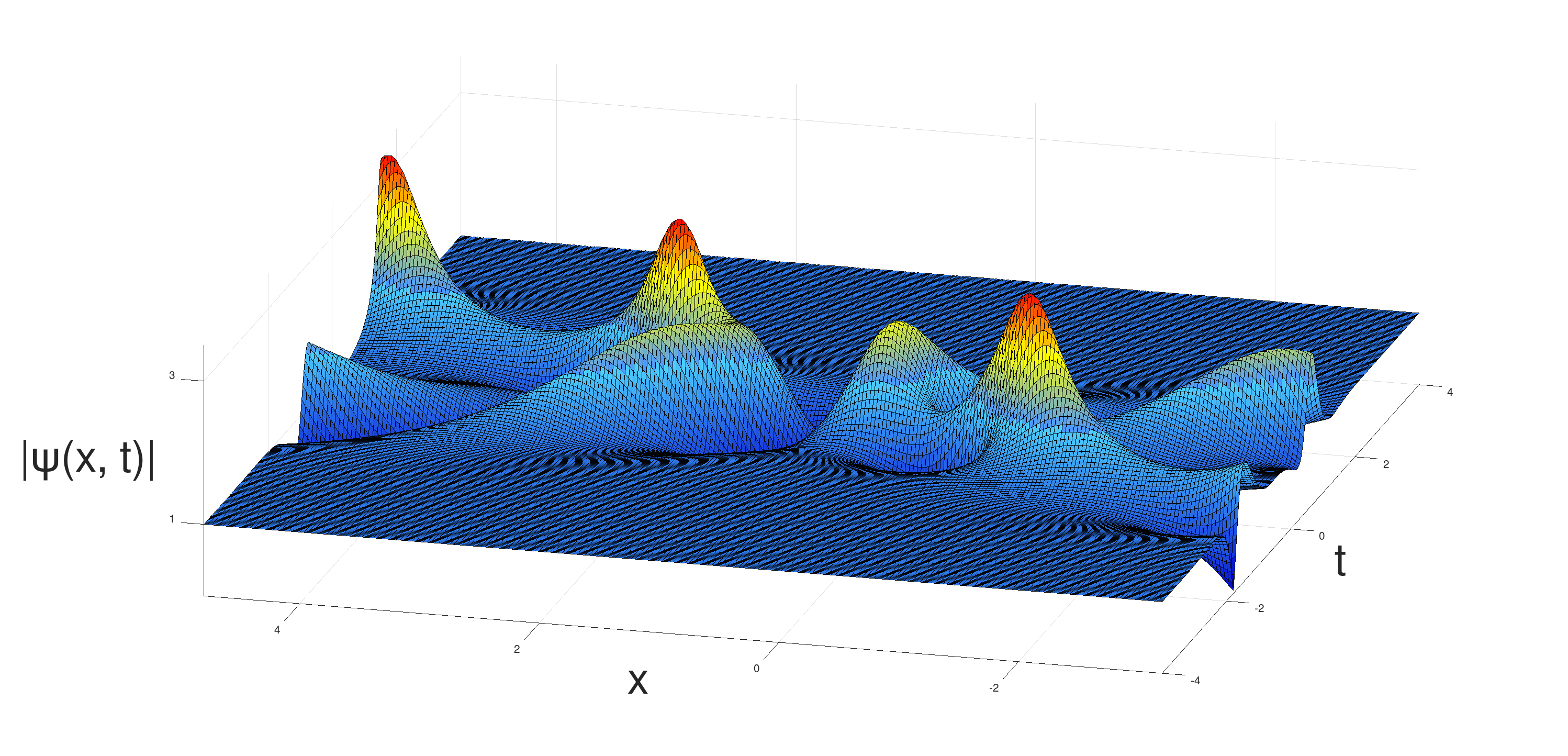}
		\centering
		\includegraphics[width=.45\linewidth]{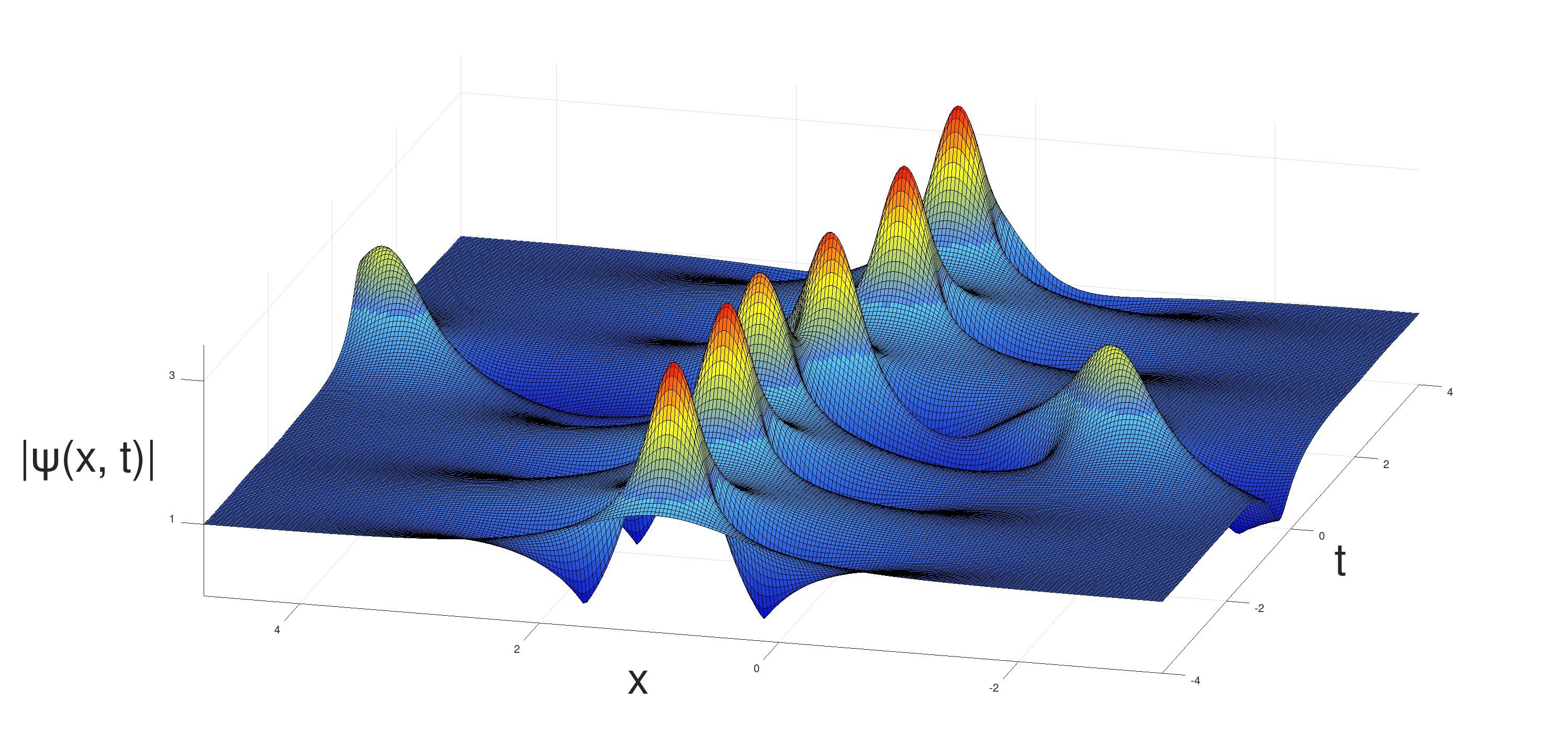}
	\caption{The three possible kinds of 2-breather solutions.Top left: a  KM-type  2-breather  with $\zeta_1=2\im$ and  $\zeta_2=3\im$.  Top right:  a TW-type 2-breather with $\zeta_1=1+2\im$ and  $\zeta_2=-2+\im$. Bottom: a 2-breather of mixed type with  $\zeta_1=2\im$ and  $\zeta_2=1+\im$.  In all three plots we set $C_1=1$ and $C_2=2$.}
	\label{fig:2breather}
\end{figure}

\noindent The Tajiri-Watanabe breather  is a non-stationary breather, since the peak does not remain localized at a fixed value in $x$ but rather moves with a speed  given by equation $\chi(x,t)=0$.
\begin{figure}[h!]
		\centering
		\includegraphics[width=.46\linewidth]{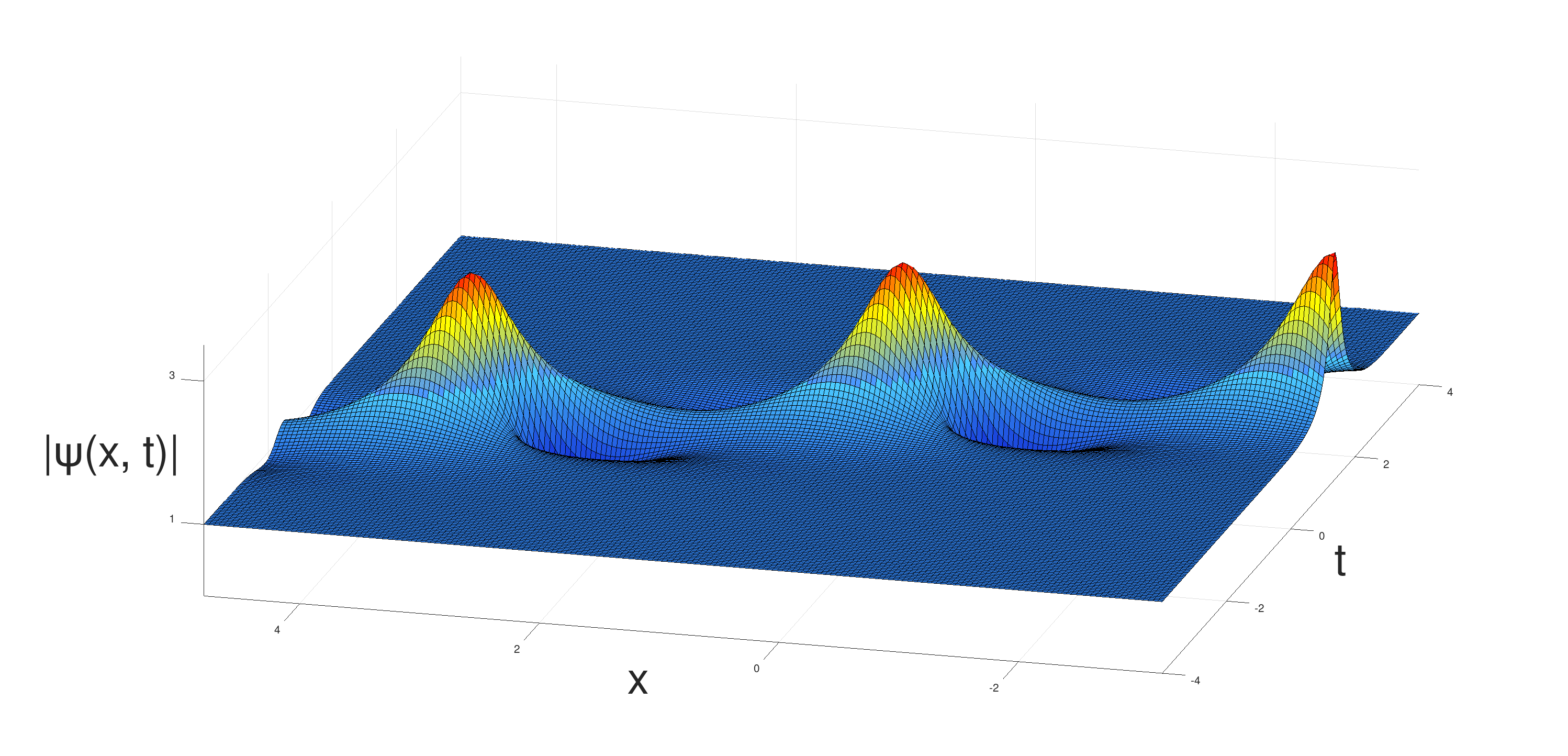}
		\centering
		\includegraphics[width=.46\linewidth]{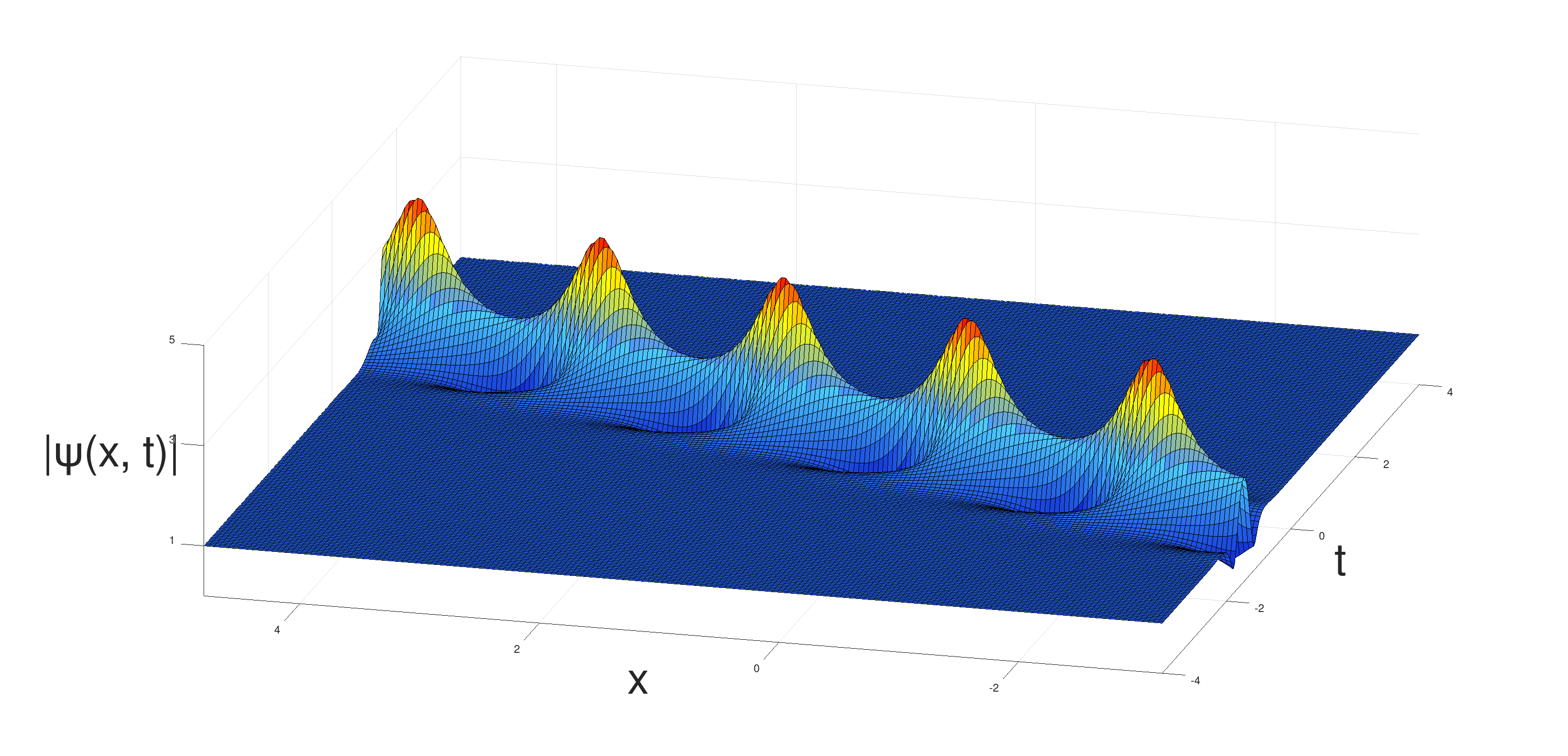}
	\caption{Two different Tajiri-Watanabe breathers, left $Z=2$, $\kappa=0$, $\phi=0$ and $\alpha=\pi/6$ and right $Z=3$, $\kappa=0$, $\phi=0$ and $\alpha=-\pi/8$.}
\end{figure}
\newline
\noindent
We have limited ourselves to display formulas for $1$-breather solutions. There exist $N$-breather solutions for every natural $N$ which can be obtained  from \eqref{Nbreathers} by  solving the linear system \eqref{A_N_breather}.  We report in Figure \ref{fig:2breather} the plot of three different 2-breather solutions obtained via different  choices of pole locations.\\

{As it is well known in the theory  of integrable systems, the  2-breather  solution   is not the sum of two 1-breather  solutions.  Such decomposition is true  for large times $t\to\pm \infty $  when the  2-breather  solution  is asymptotically close to  the sum of two 1-breather solutions,   plus o(1) \cite{Li2018}. }
\begin{remark}
In the limit $Z\to 1$ the  Tajiri-Watanbe  breather converges to the Akhmediev breather 
\begin{equation}\label{biondini twA 4.5}
		\psi_{1A}(x,t)=\frac{\cosh (2t \sin(2\alpha) + c_0' + \kappa -2\im \alpha) -\cos(\alpha)\cos(2 x \sin(\alpha) +\phi-\alpha)}{e^{2\im \alpha}(\cosh (2t \sin(2\alpha) + c_0' + \kappa) -\cos(\alpha)\cos(2 x \sin(\alpha) +\phi-\alpha))},
\end{equation}
with
\begin{equation}
	\begin{gathered}	 
		 c_0'=-\ln\left(2\cos(\alpha)|\sin(\alpha)|\right)	\,.		
 \end{gathered}
\end{equation}
In the further limit $\alpha\to 0$ one recovers the Peregrine breather.
\end{remark}
		\section{Gas of breathers and shielding} 
	\noindent	This section contains our main result. We consider a $N$-breather solution with  the first $N$ poles  $\zeta_1,\dots,\zeta_N$   filling  uniformly a ``parent" bounded  domain  $\mathcal{D}_1$ in   $D^+_1$, {with the corresponding norming constants  interpolated by a smooth function and rescaled by $N$}, and we let $N$ go to infinity. {Moreover}, we assume that $\mathcal{D}_1$ has a smooth boundary. In order to perform the  limit   $N\to\infty$	on the inverse problem for $M(z;x,t)$, we transform the residue conditions \eqref{biondini 3.8} into  jump conditions for a suitably defined Riemann Hilbert problem.   \\
	For the purpose we introduce a  close anticlockwise oriented contour  $\gamma^+_1$ in $D^+_1$  that encircles $\mathcal{D}_1$ and  we call {$\Gamma^+_1$} the  domain with boundary $\gamma^+_1$. 
	The other domains are obtained by symmetry:  in particular, $\mathcal{D}_2$  is  the domain obtained from $\mathcal{D}_1$ by the relation
\begin{equation}\label{relazione D1 D2}
				z\in \mathcal{D}_1 \Longleftrightarrow -\frac{1}{\overline{z}}\in \mathcal{D}_2\, , 
			\end{equation}
		while  $\overline{\mathcal{D}}_1$ and  $\overline{\mathcal{D}}_2$  in $D^-$ are their complex conjugates.
		
				\begin{figure}[h!]
			\centering
			\includegraphics[width=7cm]{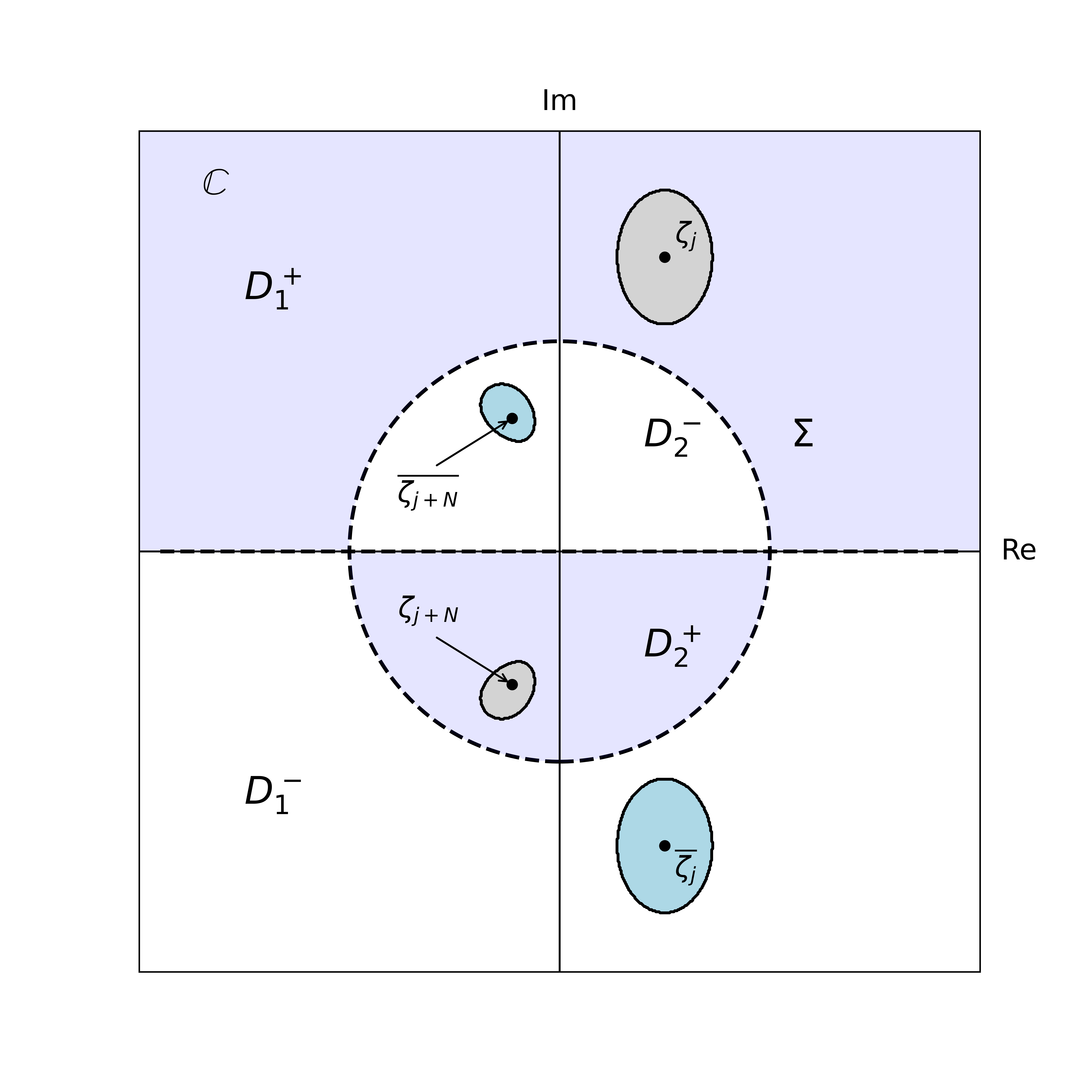}
			\includegraphics[width=7cm]{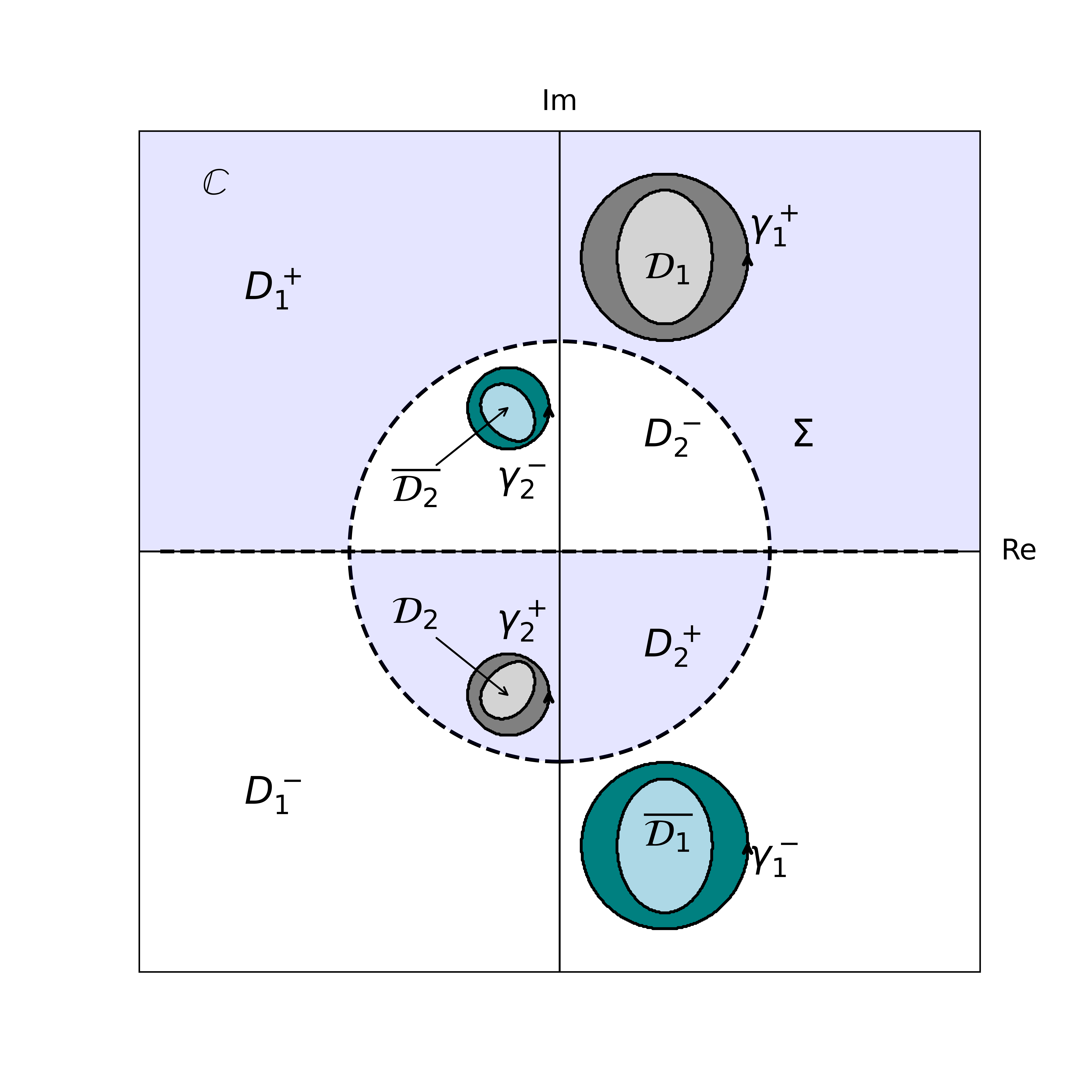}
			\caption{The complex plane with the four domains $\mathcal{D}_1$, $\mathcal{D}_2$,  $\overline{\mathcal{D}_1}$, $\overline{\mathcal{D}_2}$  and the four contours $\gamma^+_1$, $\gamma^+_2$, $\gamma^-_1$, $\gamma^-_2$ }
			\label{gamma-contours}
			\end{figure}
%
%
%
%
%
			We introduce a  closed anticlockwise oriented contour $\gamma^+_2$  that encircles $ \mathcal{D}_2$  in $ D^+_2$
		 and  we  define as  {$\Gamma^+_2$} the finite domain with boundary $\gamma^+_2$.\\
			Analogously we  define $\gamma^-_1=-\overline{\gamma^+_1}$ and $\gamma^-_2=-\overline{\gamma^+_2}$ as the closed anticlockwise oriented contours that encircle   $\overline{\mathcal{D}_1}$ and $\overline{\mathcal{D}_2}$    respectively.
			 Also in this case  we call  {$\Gamma^-_1$} and {$\Gamma^-_2$} the domains whose boundaries are $\gamma^-_1$ and $\gamma^-_2$ respectively (see Figure \ref{gamma-contours} for the geometry of these contours).\\
	We reformulate the {Riemann-Hilbert problem A} as a Riemann-Hilbert problem for a matrix	${Y}^N(z;x,t)$ that is obtained from 
	$M(z;x,t)$ by removing the poles but paying the price of putting a jump.\\	
	{More precisely,  we define the   matrix ${Y}^N(z;x,t)$  	as}		
			\begin{equation}\label{BGO relation breather}
				{Y}^N(z;x,t)=\begin{cases}
					M(z;x,t) & \text{for}\ z \in \mathbb{C}\setminus\left\{  \Gamma^+_1\cup\Gamma^+_2\cup\Gamma^-_1\cup\Gamma^-_2\right\},\\
					M(z;x,t){J}^N(z;x,t) & \text{for}\ z\in   \Gamma^+_1\cup\Gamma^+_2\cup\Gamma^-_1\cup\Gamma^-_2,
				\end{cases}
			\end{equation}
		
			{with  the jump matrix ${J}^N(z;x,t)$   having the following expression}
			\begin{equation} \label{Jt breather}
				{J}^N(z;x,t)=\left\{
				\begin{array}{ll}
					\begin{bmatrix}
						1 & 0\\
						-{ {  e}^{-2\im\theta(z)}\sum\limits_{j=1}^{N} \frac{C_{j}}{z-\zeta_{j}}}& 1
					\end{bmatrix} &\text{for}\ z\in \Gamma^+_1,\\
					\begin{bmatrix}
						1 & 0\\
						-{  e}^{2\im{\theta}(z)}{ \sum\limits_{j=N+1}^{2N} \frac{C_{j}}{z-\zeta_{j}}}& 1
					\end{bmatrix} &\text{for}\ z\in \Gamma^+_2,\\
					\begin{bmatrix}
						1 &    {  e}^{2\im \theta(z)}{\sum\limits_{j=1}^{N} \frac{\overline{C}_{j}}{z-\overline{\zeta_j}}}\\
						0& 1
					\end{bmatrix}&\text{for}\ z\in \Gamma^-_1,\\
					\begin{bmatrix}
						1 &    {  e}^{2\im \theta(z)}{\sum\limits_{j=N+1}^{2N} \frac{\overline{C}_{j}}{z-\overline{\zeta_j}}}\\
						0& 1
					\end{bmatrix}&\text{for}\ z\in \Gamma^-_2 , 
				\end{array}
				\right.
			\end{equation}
where we used the shorthand notation $\theta(z) =\theta(z;x,t) $.

			It is immediate to check that the requirement of ${Y}^N(z;x,t)$ to be analytic in {$\C\backslash \{	\boldsymbol{\gamma}\cup \{0\}\} $,} where {$
		\boldsymbol{\gamma}=\{\gamma^+_1\cup \gamma^+_2\cup\gamma^-_1\cup \gamma^-_2\} 
			$,
		  is equivalent to the residue conditions \eqref{biondini 3.8}. }\\ \bigskip
			\noindent
			Therefore, we obtain the following Riemann-Hilbert problem for the matrix function ${Y}^N(z;x,t)$. \\ \bigskip
			{\textbf{Riemann-Hilbert problem B (jump $N$-breather problem):} \\ \medskip To find a $2\times2$ matrix ${Y}^N(z;x,t)$ with the following properties: \begin{itemize}
\item  ${Y}^N(z;x,t)$ is analytic in 
$\C\backslash \{	\boldsymbol{\gamma}\cup \{0\} \}$. 			
			 \item  ${Y}^N(z;x,t)$ satisfies to the jump condition
			\begin{equation}\label{eq:R-H_sol breather }     {Y}_+^N(z;x,t)={Y}^N_-(z;x,t){J}^N(z;x,t)\end{equation}				for $z\in	\boldsymbol{\gamma}$ .  The subscript  $Y_\pm(z)$  denotes the left/right boundary values of $Y(z)$ as {$z\to	\boldsymbol{\gamma}$} in a non tangential direction.
			\item  ${Y}^N(z;x,t)$  fulfills the asymptotic conditions
			\begin{equation}
			\begin{split}
			\label{asymptoticY1}
				&  {Y}^N(z;x,t)=\mathbb{I} + \mathcal{O}\left(\frac{1}{z}\right), \quad\text{as}\ z\rightarrow \infty,\\
				&  {Y}^N(z;x,t)=\frac{\dot{\imath}}{z}\sigma_1  + \mathcal{O}\left(1\right), \quad\text{as}\ z\rightarrow 0\,  .
				\end{split}
			\end{equation}
			\end{itemize} 		
		}
\noindent	Finally, the $N$-breather solution of the focusing nonlinear Schr\"odinger equation is recovered  from the matrix  ${Y}^N(z;x,t)$
	 by the relation
		\begin{equation}\label{Nbreathers tilde}
		\psi(x,t)=-\im \lim_{z\rightarrow \infty} z {Y}_{12}^N(z;x,t)\,.
	\end{equation}
		\subsection{Gas of breathers}
Now we have reformulated the inverse problem in such a way that inspecting the jump matrix \eqref{Jt breather}, it is possible to perform the limit $N\to\infty$ by assuming that the point spectrum $\zeta_1,\dots \zeta_N$ of the breathers    accumulates  uniformly inside the  domain 	 { $\mathcal{D}_1\subset \Gamma^+_1$} as $N\to\infty$. 
			\begin{figure}[h!]
			\centering
			\includegraphics[width=7cm]{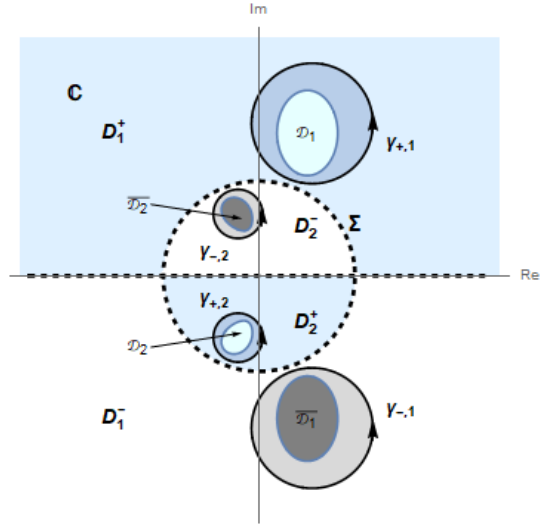}
			\caption{The complex plane with the four domains $\mathcal{D}_1$, $\mathcal{D}_2$,  $\overline{\mathcal{D}_1}$, $\overline{\mathcal{D}_2}$ }
			\label{piano complesso con cerchi tutto}
			\label{gamma-contours1}
			\end{figure}
			  Namely,  we assume that  the spectral breather density  converges to the uniform measure of the domain $\mathcal{D}_1$, that is 	
		\begin{equation}
		\label{uniform1}
		\frac{\mathcal{A}_1}{N}\overset{N\to\infty}{\to} d^2w,
		\end{equation}
		where $\mathcal{A}_1$ is the area  of  $\mathcal{D}_1$ and $d^2w$ is the infinitesimal cartesian area.

			We interpolate the norming constants as
			\begin{equation}\label{cj breathers}
				\begin{aligned}
					&C_j=\frac{\mathcal{A}_1}{\pi N}\beta_1\left(\zeta_j,\overline{\zeta_j}\right), \\
					& C_{N+j}=\frac{\mathcal{A}_1}{\pi N}\beta_2\left(\zeta_{N+j},\overline{\zeta_{N+j}}\right), 
				\end{aligned}
			\quad j=1,...,N\, , 
			\end{equation}
			where $\beta_1(\cdot,\cdot)$ and  $\beta_2(\cdot,\cdot)$ are  smooth functions, and we  use the notation $\beta_1\left(\zeta_j,\overline{\zeta_j}\right)$ to stress the fact that the function $\beta_1$ is not analytic. Since   the norming constants satisfy the conditions \eqref{biondini norming constants 1} and \eqref{biondini norming constants 2}, in particular 			
			\begin{equation}\label{symm prop}
							C_{N+j}=-\left(\frac{1}{\overline{\zeta_j}}\right)^2 \overline{C_j}, \quad j=1,...,N,
						\end{equation}
						it follows that 	
						\[
						\beta_2\left(\zeta_{N+j},\overline{\zeta_{N+j}}\right)	=-\left(\frac{1}{\overline{\zeta_j}}\right)^2\overline{\beta_1\left(\zeta_j,\overline{\zeta_j}\right)}\]
						or, equivalently		using the relation $	\zeta_{N+j}=-\frac{1}{\overline{\zeta_j}}$, 
						\begin{equation}
						\beta_2\left(z,\overline{z}\right)=-z^2\overline{\beta_1\left(-\frac{1}{\overline{z}}, -\frac{1}{{z}}\right)},\quad z\in \mathcal{D}_2.
					\end{equation}
		In this way the sum in  the off-diagonal entries of  the jump matrix ${J}^N$ in  \eqref{Jt breather} becomes a Riemann sum and we have 		\small
			\begin{equation}\label{BGO monte carlo breathers 00}
				\begin{gathered}
				\sum_{j=1}^{N}\frac{C_j }{(z-\zeta_j)}=\sum_{j=1}^{N}\frac{\mathcal{A}_1}{\pi N} \frac{\beta_1(\zeta_j,\overline{\zeta_j})}{(z-\zeta_j)}
				\mathop{\longrightarrow}_{N\to\infty} \iint_{\mathcal{D}_1}\frac{\beta_1(w,\overline{w}) }{z-w}\frac{d^2w}{\pi},\\
				\sum_{j=N+1}^{2N}\frac{C_j }{(z-\zeta_j)}=\sum_{j=N+1}^{2N}\frac{\mathcal{A}_1}{\pi N} \frac{\beta_2(\zeta_j,\overline{\zeta_j}) 
				}{z-\zeta_j} \mathop{\longrightarrow}_{N\to\infty} \iint_{\mathcal{D}_2}\frac{\beta_2(w,\overline{w})}{z-w}\frac{d^2w}{\overline{w}^2w^2\pi}\,, 
			\end{gathered}
			\end{equation}
			{  where  the second relation  follows from the fact the the uniform area measure in  $\mathcal{D}_1$  becomes, under  the coordinate transformation   $w\to -\dfrac{1}{\overline{w}}$, the measure $\dfrac{d^2w}{\overline{w}^2w^2}$ in $\mathcal{D}_2$. }

		\normalsize
			Since the ``theta" condition 
						\begin{equation}\label{theta}
							-\arg\left({\psi_+}\right) = 4\sum_{n=1}^N\arg\left(\zeta_n\right)    
						\end{equation}
						must hold, we assume that the domain $\mathcal{D}_1$ (and  consequentially  $\overline{\mathcal{D}_1}$,  $\mathcal{D}_2$  and $\overline{\mathcal{D}_2}$) is chosen such that \eqref{theta} holds also when passing to the limit $N\rightarrow\infty$. \\\bigskip		{	Therefore  the Riemann-Hilbert problem B, when $N\rightarrow\infty$, transforms into the following .\\ \bigskip 
		\textbf{Riemann-Hilbert problem C (jump $N\to\infty$ problem):} \\ \medskip To seek for a $2\times2$ matrix ${Y}^\infty(z;x,t)$ with the following properties:
		\begin{itemize}
			\item  ${Y}^\infty(z;x,t)$ is analytic in 
			$\C\backslash \{	\boldsymbol{\gamma}\cup \{0\} \}$.
			\item  ${Y}^\infty(z;x,t)$ fulfills the jump condition 
			$\C\backslash \{	\boldsymbol{\gamma}\cup \{0\} \}$.
				\begin{equation}\label{eq:R-H_sol2 breather jump}
			{Y}_+^\infty(z;x,t)={Y}_-^\infty(z;x,t)( \mathbb{I}+{J}^\infty(z;x,t)),\quad z\in\boldsymbol{\gamma},
			\end{equation}
			with
		\begin{equation}\label{RH 3 Jump}
			\begin{split}
				&	{J}^\infty(z;x,t)=							
				\left[\begin{matrix}
					0 & 	e^{2\im \theta(z)} {J}_{12}^\infty(z;x,t)	\\
					-e^{-2\im \theta(z)}{J}_{21}^\infty(z;x,t) & 0
				\end{matrix}\right],\\
				&	{J}_{21}^\infty(z;x,t)=	\iint\limits_{\mathcal{D}_1}\frac{ \beta_1(w,\overline{w}) d^2w}{\pi(w-z)}{\bf 1}_{\gamma^+_1}(z)+\iint\limits_{\mathcal{D}_2}\frac{ \beta_2(w,\overline{w}) d^2w}{\pi(w-z)\overline{w}^2w^2}{\bf 1}_{\gamma^+_2}(z)\,,\\
				&	{J}_{12}^\infty(z;x,t)=	\iint\limits_{\overline{\mathcal{D}_1}}\frac{ \beta_1^{*}(w,\overline{w}) d^2w}{\pi(w-z)}{\bf 1}_{\gamma^-_1}(z) +  \iint\limits_{\overline{\mathcal{D}_2}}\frac{  \beta_2^{*}(w,\overline{w}) d^2w}{\pi(w-z)\overline{w}^2w^2} {\bf 1}_{\gamma^-_2}(z)\,,
			\end{split}
		\end{equation}
		where $\beta_1^{*}(w,\overline{w})= \overline{\beta_1(\overline{w},w)}$,  $\beta_2^{*}(w,\overline{w})= \overline{\beta_2(\overline{w},w)}$, and  ${\bf 1}_{\gamma}$ is the charactheristic function of the contour $\gamma$.
			\item  ${Y}^\infty(z;x,t)$ satisfies the asymptotic conditions 
		\begin{equation}\label{eq:R-H_sol2 breather asymptotics}
			\left\{	\begin{split}
					& {Y}^\infty(z;x,t)= \mathbb{I}+\mathcal{O}\left(\frac{1}{z}\right),\quad\text{as}\ z\rightarrow \infty,\\
					& {Y}^\infty(z;x,t)= \frac{\im}{z} \begin{bmatrix}
						0 &1\\
						1&0
					\end{bmatrix}+\mathcal{O}\left(1\right),\quad\text{as}\ z\rightarrow 0.
				\end{split}\right.
			\end{equation}		
				\end{itemize}	}
								\begin{theorem}
[\bf Existence of the solution $\psi_{\infty}$] \label{Theorem2.4}
There is a unique solution $ {Y}^\infty(z;x,t)$ to the limiting   Riemann-Hilbert problem C, which determines a solution $\psi_{\infty}(x,t)$ to the focusing NLS equation via 
\begin{equation}
\label{Isoliton}
\begin{split}
&\psi_{\infty}(x,t)=-\im \lim_{z \to \infty}z{Y}_{12}^\infty(z;x,t).
\end{split}
\end{equation}
Moreover, $\psi_\infty$ is a classical solution to the focusing NLS equation, which belongs to the class $C^\infty(\R\times\R^+)$.
\end{theorem}
															The proof of the above theorem is reported in Appendix~\ref{AppB}.	
															
																				While the properties of the class of solutions obtained by solving the above Riemann-Hilbert problem for $Y^\infty$ is not explored, in the section below we can get an insight  for particular choices of the domain $\mathcal{D}_1$ and the breather density $\beta_1$. Using Green's theorem and the Cauchy theorem we can solve the problem exactly and reduce it to a finite number of breathers.
						\begin{remark}
						In the case of a gas of Akhmediev breathers,  we can let the point $\zeta_1,\dots \zeta_N$ accumulate on  an arc $\mathcal{L}_1$ of the unit circle. The corresponding gas of Akhmediev breathers  is described by the solution of the Riemann-Hilbert problem {C}
				  where the jump matrix ${J}^\infty(z;x,t)$ has entries
					\begin{equation}
							\begin{split}
												&	{J}_{21}^\infty(z;x,t)=	\int\limits_{\mathcal{L}_1}\frac{ \beta_1(w,\overline{w}) dw}{\pi(w-z)}{\bf 1}_{\gamma^+_1}(z)+\int\limits_{\mathcal{L}_2}\frac{ \beta_2(w,\overline{w}) dw}{\pi(w-z)}{\bf 1}_{\gamma^+_2}(z)\,,\\
					&	{J}_{12}^\infty(z;x,t)=	\int\limits_{\overline{\mathcal{L}_1}}\frac{ \beta_1^{*}(w,\overline{w}) d^2w}{\pi(z-w)}{\bf 1}_{\gamma^-_1}(z) +  \int\limits_{\overline{\mathcal{L}_2}}\frac{  \beta_2^{*}(w,\overline{w}) dw}{\pi(z-w)} {\bf 1}_{\gamma^-_2}(z)\,,
						\end{split}
						\end{equation}	
						where 	$z\in \mathcal{L}_2$ if   $-z\in \mathcal{L}_1$.		The contours $\gamma^+_1$ and  $\gamma^+_2$	  encircle anti-clockwise the arcs	$\mathcal{L}_1$ and $\mathcal{L}_2$, respectively and similarly 		the  contours $\gamma^-_1$ and  $\gamma^-_2$	  encircle anti-clockwise the arcs	$\overline{\mathcal{L}_1}$  and $\overline{\mathcal{L}_2}$, respectively.  We assume that the arcs 	$\mathcal{L}_1$,  $\mathcal{L}_2$, $\overline{\mathcal{L}_1}$ and $\overline{\mathcal{L}_2}$ do not intersect. By a transformation  as in \eqref{BGO relation breather}, the above Riemann-Hilbert problem can be mapped to a Riemann-Hilbert problem with jumps on the contours $\mathcal{L}_1, \mathcal{L}_2, \overline{\mathcal{L}_1}$ and $\overline{\mathcal{L}_2}$.  The initial data associated to such gas of breathers
						is expected to be step-like oscillatory, namely for  $x\to -\infty$ one expects to have a travelling wave and  for $x\to +\infty$ one expect that the initial data is approaching   a  nonzero constant of modulus one.
						 This is a generalization of the soliton gas considered in \cite{BGO2024}.						\end{remark}
						\subsection{Shielding of breathers on quadrature domains}
						In this section we take a particular  class of  domains  $\mathcal{D}_1$ and interpolating function $\beta_1$ so that 
						the gas of breathers turns into a finite set of breathers.
						Let us start by considering the class of quadrature domains  \cite{Gustafsson_1983}
						\begin{equation}\label{D_1}
							\mathcal{D}_1:= \left \{ z\in \mathbb{C} \text{ s.t. }  \Big|(z-d_{0}  )^{m}-d_{1}\Big|<\rho\right\},\;\;m\in\mathbb{N},
						\end{equation}
						with $d_{0}\in D^+_1$ and  $|d_{1}|,\ \rho>0$ chosen so that $ \mathcal{D}_1\subset D^+_1$.
							The boundary of $\mathcal{D}_1$  is described by the Schwartz function \cite{Gustafsson_1983}:
						\begin{equation}
							\label{Schwartz1 D_j}
							\begin{gathered}
							\overline{z}=S_1(z),\\ S_1(z)=\overline{d_{0}}+\left(\overline{d_{1}}+\frac{\rho^2}{(z-d_{0})^{m}-d_{1}}\right)^{\frac{1}{m}}\,.
					\end{gathered}
					\end{equation}
					The domain  $\mathcal{D}_2$  is defined by symmetry:
										\begin{equation}\label{D_2}
						\mathcal{D}_2=\left \{ z\in \mathbb{C} \text{ s.t. } - \frac{1}{\overline{z}}\in \mathcal{D}_1\right\},
					\end{equation}
					that gives
					\begin{equation}\label{D_2a}
							\mathcal{D}_2:= \left \{ z\in \mathbb{C} \text{ s.t. }  \Big|(-\frac{1}{z}-\overline{d_{0}}  )^{m}-\overline{d_{1}}\Big|<\rho\right\},\;\;m\in\mathbb{N},
						\end{equation}
						with boundary given by the relation
						\begin{equation}
							\label{Schwartz1 D_2}
							\begin{gathered}
							\overline{z}=S_2(z),\quad  S_2(z)=-\frac{1}{d_0+\left(d_1+\frac{\rho^2}{(-\frac{1}{z}-\overline{d_{0}})^{m}-\overline{d_{1}}}\right)^{\frac{1}{m}}}\,.
					\end{gathered}
					\end{equation}
							Recalling that $\psi_-=\psi_0=1$, we choose the asymptotic value $\psi_+$ to be as simple as possibile,  namely $\psi_+=1$, leading to the ``theta" condition \eqref{biondini 3.14 theta cond breather}
						\begin{equation}\label{particular theta}
							4\sum_{j=1}^N\arg(\zeta_j)=0\,,\quad (\text{mod}\ 2\pi).
						\end{equation}
						This condition  is satisfied if 
					 $\zeta_j$  lies on the imaginary axis, or pairs of points $\zeta_j$ are symmetric with respect to the imaginary axis. This can be achieved by taking the domain $\mathcal{D}_1$ symmetric with respect to the imaginary axis. 
					 A possible choice to realize this condition is 
					 \begin{equation}
					 \begin{split}
					 \label{theta_cond}
&					 \Im (d_0)>1,\, \Re (d_0)=0,\Re (d_1)=0,\quad \mbox{$m$ odd},\\
& \Im (d_0)>1, \, \Re (d_0)=0,\, \Im (d_1)=0,    \quad \mbox{$m$ even}, 
 \end{split}
  \end{equation}
 where $\Re(z)$ and $\Im(z)$ are, respectively, the real and imaginary parts of teh complex number $z$.
					 We  also remark  that the constants $d_0,d_1$ and $\rho$ have to be tuned  in such a way that 
					 the domain $\mathcal{D}_1$ lies in $D_1$. 
					We chose the interpolating function $\beta_1$ of the form 
					\begin{equation}\label{beta1def}
					\beta_1(z,\overline{z})= n(\overline{z}-\overline{d_{0}})^{n-1}r(z)=\overline{\partial}\left((\overline{z}-\overline{d_{0}})^{n})r(z)\right)\, , 
					\end{equation} 
					    with  $r(z)$   analytic in $\mathcal{D}_1$. Then 
					\begin{equation}\label{d2scw}
						\beta_2(z,\overline{z})=-n\left(-\frac{1}{\overline{z}}-d_0\right)^{n-1}z^2\, \overline{r\left(-\frac{1}{\overline{z}}\right)}, \quad z\in \mathcal{D}_2
					\end{equation}
					and we note that by the Schwartz reflection  principle $\overline{r\left(-\frac{1}{\overline{z}}\right)}$ is analytic in $\mathcal{D}_2$.

						Let us insert our choices in the expression  \eqref{RH 3 Jump}  for the jump matrix ${J}^\infty$ in \eqref{RH 3 Jump}.
						 For $z\notin \mathcal{D}_1$ 
						we obtain
						\begin{equation}
							 \label{eq:int_prob1 breathers}
							\iint\limits_{\mathcal{D}_1}\frac{\beta_1(w,\overline{w})}{\pi(z-w)} d^2w=\frac{1}{2\pi \im} \oint_{\partial\mathcal{D}_1}\frac{r(w)(\overline{w}-\overline{d_{0}})^{n}}{ z-w}\ dw\\							=\frac{ 1}{2\pi \im } \oint_{\partial \mathcal{D}_1 }\frac{
							(S_1(w)-\overline{d_{0}})^n r(w)}{z- w}\ dw
													\end{equation}
													where $S_1$ is the Schwartz function \eqref{Schwartz1 D_j}. {The first equality in \eqref{eq:int_prob1 breathers} is due to  Green's theorem in its complex form, while  in the second one we use the fact that the boundary $\partial \mathcal{D}_1$ of $ \mathcal{D}_1$ is described by $\overline{z}=S_1(z)$}. 
													
													{ The next  step  is to consider the case $m=n$ so that the integral  \eqref{eq:int_prob1 breathers} can be evaluated using  the  residue theorem, thus 
							 showing that a $n$-breather  solution  emerges via this relation between the
							 the quadrature domain defined by formulas \eqref{D_1} and \eqref{Schwartz1 D_j} and the interpolating function.}\\   
						Indeed   using the residue theorem to  the contour integral \eqref{eq:int_prob1 breathers} at the  $n$  poles  $\{\lambda_{1},\dots, \lambda_{n}\}$ 
						given by the solution of the equation $(z-d_{0})^n=d_{1}$, it follows that: 
						\begin{equation}\label{greenShielding}
							\begin{split}    
								\iint\limits_{\mathcal{D}_1}\frac{\beta_1(w,\overline{w})}{\pi(z-w)}  d^2w
								&=\frac{ 1}{2\pi \im } \oint_{\partial \mathcal{D}_1 }\frac{(S_1(w)-\overline{d_{0}})^n r(w)}{z- w}\ dw\\
								&= \rho^2\sum_{j=1}^{n}\frac{r(\lambda_{j})}{\prod_{k\ne j}(\lambda_{j} - \lambda_{k})}\dfrac{1}{(z-\lambda_{j})},   \quad \text{for $z\notin \mathcal{D}_1$}  \,. 
							\end{split}
						\end{equation}	
						{ Comparing the above relation with jump matrix  \eqref{Jt breather} for the $N$-breather solution 
						yields  the norming constants }
						\begin{equation}\label{cjshielding}
							C_j=\frac{\rho^2 r(\lambda_{j})}{\prod_{k\ne j}(\lambda_{j} - \lambda_{k})}\,.
						\end{equation}

							{In particular, it is evident that the emergence of the $n$-breather solution is due to the   term $n(\overline{z}-\overline{d_{0}})^{n-1}$  in \eqref{beta1def}, while the  function $r(z)$  influences the value of the norming constants, due to \eqref{cjshielding}, hence contributing only to the ``shape" of the $n$-breather. Moreover, we remark that the first equality in \eqref{greenShielding} does  not depend on the choice $m=n$. On the other hand, the choice  $m=n$ ensured the validity of the second equality of \eqref{greenShielding} holds.  }\\	
				Since the  $n$  poles  $\{\lambda_{1},\dots, \lambda_{n}\}$ 
				are the solution of the equation $(z-d_{0})^n=d_{1}$, they have the form
			\begin{equation}\label{lambdak}
				\lambda_{k}=d_0+\sqrt[n]{|d_1|}e^{\im\left(\frac{\arg(d_1)}{n}+\frac{2(k-1)\pi}{n}\right)}\, \;\text{for}\ k=1,...,n\, , 
					\end{equation}
			and hence from \eqref{theta_cond}
						the ``theta" condition
					$4\sum_{j=1}^n\arg(\lambda_j)=0,\, \mathrm{mod } \, 2\pi$ is satisfied.

						The integral over $\mathcal{D}_2$ gives 
						a result compatible with the symmetry constraints. Indeed we have, also taking \eqref{d2scw} into account, 
						\begin{equation}\label{AltriC}
						\begin{aligned}
						&\iint\limits_{\mathcal{D}_2}\frac{ \beta_2(w,\overline{w}) d^2w}{\pi(w-z)\overline{w}^2w^2}=-n\iint\limits_{\mathcal{D}_2}\frac{\left(-\frac{1}{\overline{w}}-d_0\right)^{n-1}w^2\, \overline{r\left(-\frac{1}{\overline{w}}\right)} d^2w}{\pi(w-z)\overline{w}^2w^2}\\
						&=-\int\limits_{\partial\mathcal{D}_2}\frac{\left(-\frac{1}{\overline{w}}-d_0\right)^{n}\, \overline{r\left(-\frac{1}{\overline{w}}\right)} dw}{2 i\pi(w-z)}=\int\limits_{\partial\mathcal{D}_2}\frac{\left(-\frac{1}{S_2(w)}-d_0\right)^{n}\, \overline{r\left(-\frac{1}{\overline{w}}\right)} dw}{2 i\pi(w-z)}\\
=&-\int\limits_{\partial\mathcal{D}_2}\frac{\left(d_1+\frac{\rho^2}{(-\frac{1}{w}-\overline{d_{0}})^{n}-\overline{d_{1}}}\right)\, \overline{r\left(-\frac{1}{\overline{w}}\right)} dw}{2 i\pi(w-z)}\\		
						&=-\frac{ \rho^2}{(-\overline{d_0})^n-\overline{d_1}}\sum_{j=1}^{n}\frac{\lambda_{j+n}^n\overline{  r(-\frac{1}{\overline{\lambda_{j+n}}})}}{\prod_{k\ne j}(\lambda_{j+n} - \lambda_{k+n})}\dfrac{1}{(z-\lambda_{j+n})}\, .						\end{aligned}
						\end{equation}
						{ In this equation  $\lambda_{n+j}$, $j=1,\dots, n$ are the $n$ zeros of the equation $(-\frac{1}{z}-\overline{d_{0}})^{n}-\overline{d_{1}}=0$\,,. We remark that the second equality holds thanks to the fact that $\overline{r\left(-\dsl{\frac{1}{\overline{z}}}\right)}$ is analytic because of  Schwartz reflection  principle.}
						
						 From \eqref{AltriC} it is immediate to verify that 
												\begin{equation}
							\lambda_{n+j}=-\frac{1}{\overline{\lambda_{j}}},\quad j=1,...,n\,,
						\end{equation}
						and  the  norming constants $C_{n+j}$ given by 
						\begin{equation}
						C_{n+j}=-\frac{ \rho^2}{(-\overline{d_0})^n-\overline{d_1}}\lambda_{j+n}^{n}\frac{\overline{r(-\frac{1}{\overline{\lambda_{j+n}}})}}{\prod_{k\ne j}(\lambda_{j+n} - \lambda_{k+n})}
						\end{equation}
						satisfy the symmetry 
						\begin{equation}\label{neq3}
							C_{n+j}=-\left(\frac{1}{\overline{\lambda_{j}}}\right)^2\overline{C_j},\quad j=1,...,n\, .
						\end{equation}  
						Adding up the two integrals over $\mathcal{D}_1$ and $\mathcal{D}_2$, we obtain, up to a sign, the entry $2,1$ of the jump matrix \eqref{Jt breather}.\\
						Therefore we can conclude that the  breather gas solution $\psi_{\infty}(x,t)$ in \eqref{Isoliton} coincides with the $n$-breather solution  $\psi_n(x,t)$ in \eqref{Nbreathers tilde} with spectrum $\{\lambda_{1},\dots,\lambda_{2n}\}$, with $\lambda_{n+j}=-\frac{1}{\overline{\lambda_{j}}}$, for $j=1,..., n$, and corresponding 
						norming constants given by \eqref{cjshielding} and \eqref{neq3}.
						We summarize the section with the following theorem
						\begin{theorem}
						The gas breather solution \eqref{Isoliton}   obtained from the Riemann-Hilbert problem  {C} is reduced to a $n$-breather solution  when the  domain $\mathcal{D}_1$   is a quadrature domain of the form
						\[
						\mathcal{D}_1:= \left \{ z\in \mathbb{C} \text{ s.t. }  \Big|(z-d_{0}  )^{n}-d_{1}\Big|<\rho\right\},\;\;n\in\mathbb{N},
\]
where the constants $d_0$, $d_1$, $\rho$ and $n$ are subject to the constraints \eqref{theta_cond} and  tuned in such a way that $\mathcal{D}_1\subset D_1$  and the interpolating function   is  of the form
\[
\beta_1(z,\overline{z})=n(\overline{z}-\overline{d_{0}})^{n-1}r(z)\,
\]
where $r(z)$ is analytic in $\mathcal{D}_1$.
The spectrum of the $n$-breather solution is given by the zeros  $\{\lambda_{1},\dots, \lambda_{n}\}$  
of the polynomial $(z-d_{0}  )^{n}-d_{1}=0$ and the norming constants
are 
\[
C_j=\frac{\rho^2 r(\lambda_{j})}{\prod_{k\ne j}(\lambda_{j} - \lambda_{k})}\,.
\]

						\end{theorem}

				\subsection{Examples}
				We end this Section by considering the simplest instances of quadrature domains leading to effective $n=1$ and $n=2$ breathers. 
						
				\subsubsection{The $n=1$ case: the Kutznesov-Ma breather}
						\noindent
						In the particular case $n=1$, $\mathcal{D}_1$ is the disk of radius $\rho$ centered at $d_{0}+d_1=\im\varsigma$  with $\varsigma>1$
							\begin{equation}\label{D_1-2}
							\mathcal{D}_1:= \left \{ z\in \mathbb{C} \text{ s.t. }  \Big|z-\im\varsigma\Big|<\rho\right \},
						\end{equation}   whose boundary is described by the Schwarz function
					\begin{equation}
						\label{Schwartz circle}
						\overline{z}=S_{disk}(z),\quad S_{disk}(z)=-\im\varsigma+\frac{\rho^2}{(z-\im\varsigma)}.
					\end{equation}
					 We obtain  exactly the  Riemann-Hilbert problem \eqref{eq:R-H_sol breather }, \eqref{asymptoticY1} for $N=1$ with $\mathcal{Z}=\{\lambda_{1}=\im\varsigma,\ \lambda_{2}=-\im/\varsigma\}$, $C_1=\rho^2r(\im\varsigma)$ and $C_2=-\left(\frac{i}{\varsigma}\right)^2\overline{\rho^2r(\im\varsigma)}$.\\
					 Writing $C_1=e^{\kappa_0 + \im \phi_0}$ we obtain
					\begin{equation}\label{biondini 4.1 shielding}
						\psi_1(x,t)=\frac{\cosh{(\widetilde{\chi})}-\frac{1}{2}\widetilde{c_+}\left(1+\frac{\widetilde{c_-}^2}{\widetilde{c_+}^2}\right)\sin{(\widetilde{s})}+\im\widetilde{c_-}\cos{(\widetilde{s})}}{\cosh{(\widetilde{\chi})}-A\sin{(\widetilde{s})}}
					\end{equation}
					with
					\begin{equation}\label{biondini 4.2 shielding}
						\begin{gathered}						
						A=\frac{2}{\widetilde{c_+}}<1, \quad
						\widetilde{c_{\pm}}=\varsigma\pm\frac{1}{\varsigma}, \\ 
						\widetilde{\chi}(x,t)=\widetilde{c_-}x+\widetilde{c_0}+\kappa_0, \quad  \widetilde{c_0}=			 \ln\left(\frac{\widetilde{c_+}}{2\varsigma\widetilde{c_-}}\right),\\
						 s(x,t)=\widetilde{c_+}\widetilde{c_-}t-\phi_0.
							\end{gathered}
					\end{equation}
			 \begin{figure}[h!]
					\centering
					\includegraphics[width=.3\linewidth]{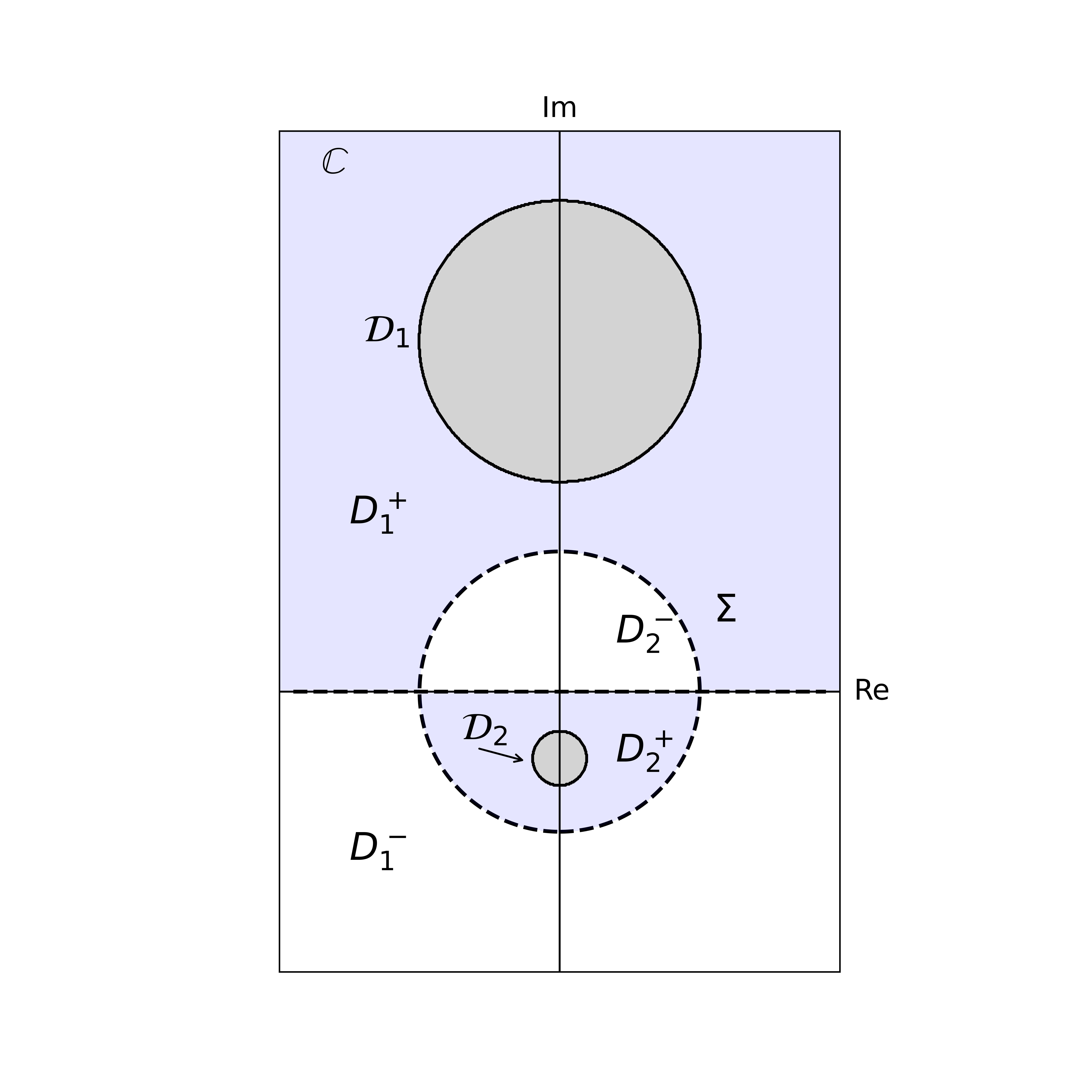}
					\includegraphics[width=.5\linewidth]{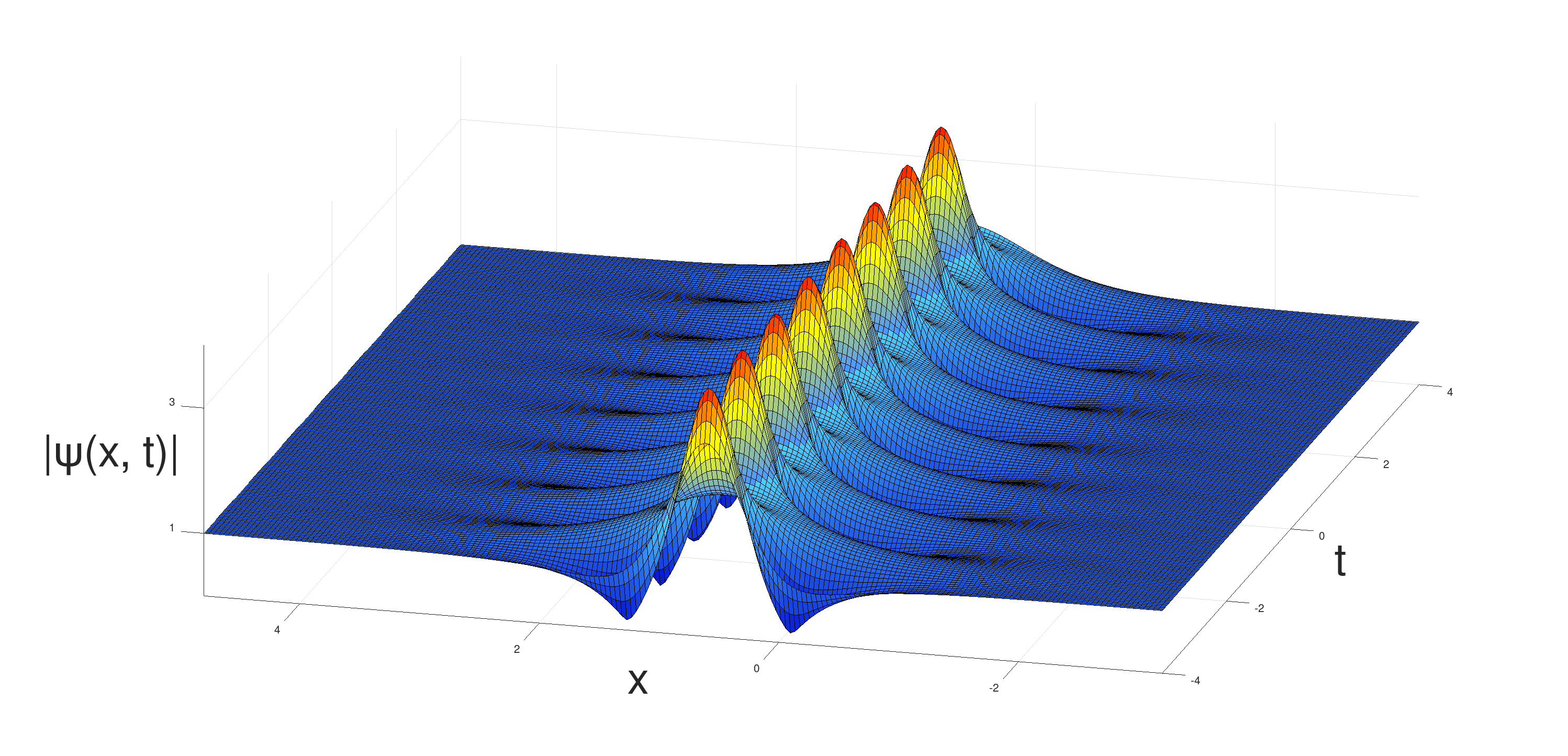}
				\caption{Left: The 1-fold domain of accumulation: a disk of radius 
				 $\rho=1$ centred at $\frac{5}{2}\im$. Right: the ``shielding" Kuznetsov-Ma breather. Parameters: $\lambda_1=\frac{5}{2}\im$, $C_1=1$, $r(z)=1$.}
				\label{KM1eff}
			\end{figure}

				\subsubsection{The $n=2$ case: the Tajiri-Watanabe  and Kutznesov-Ma   breathers}
\begin{figure}[h!]
						\centering
						\includegraphics[width=.3\linewidth]{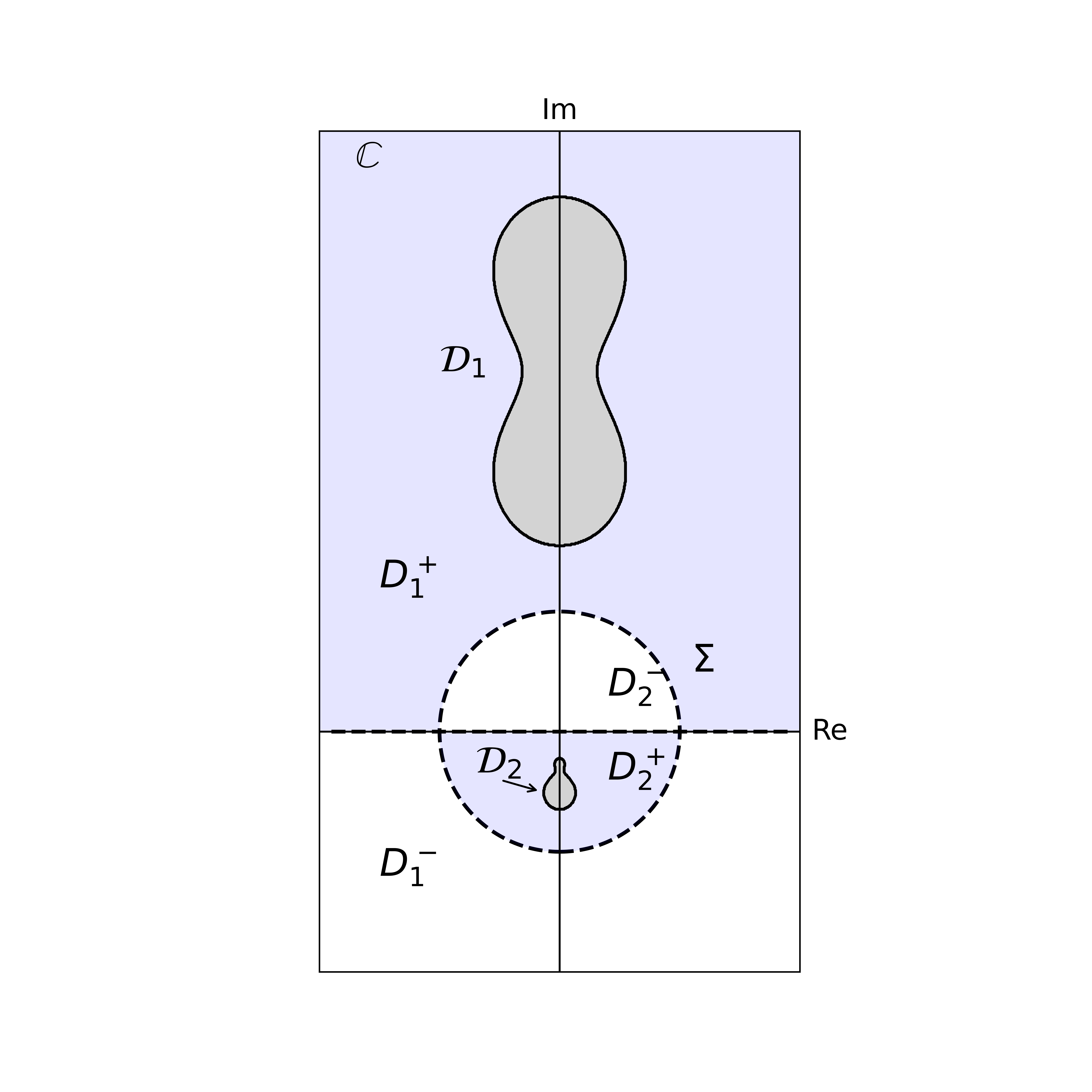}
							\includegraphics[width=.3\linewidth]{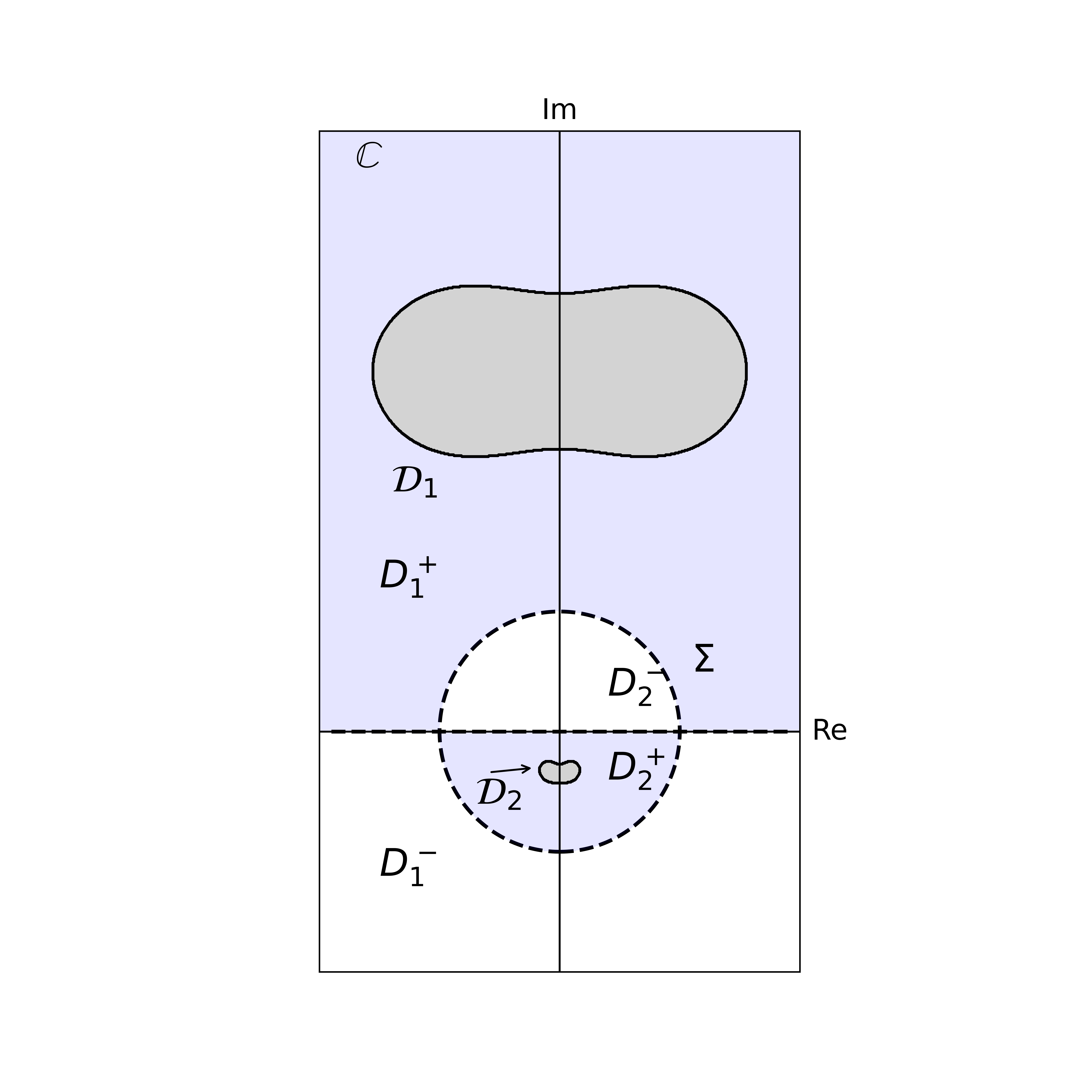}
%
						\centering
						\includegraphics[width=.45\linewidth]{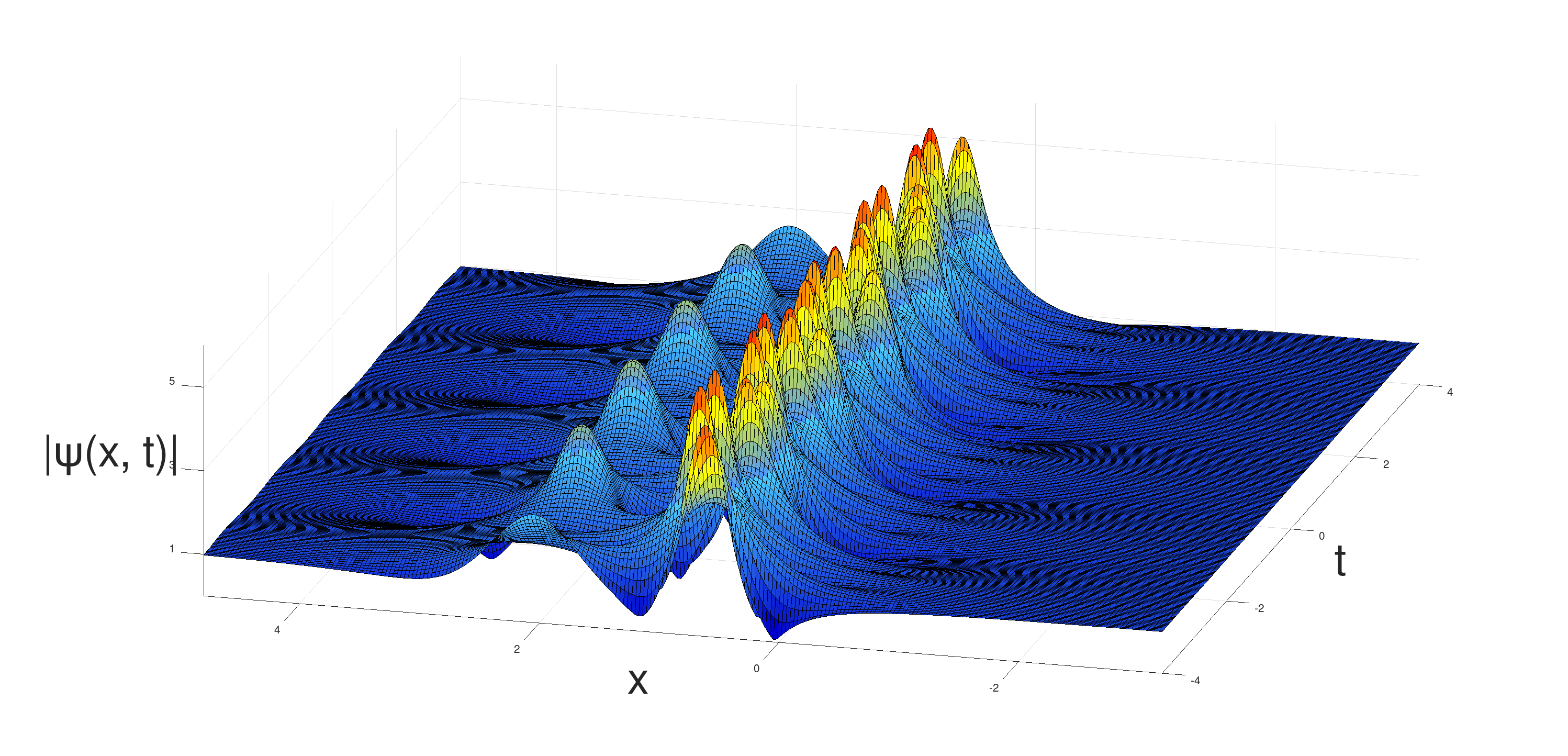}
						\includegraphics[width=.45\linewidth]{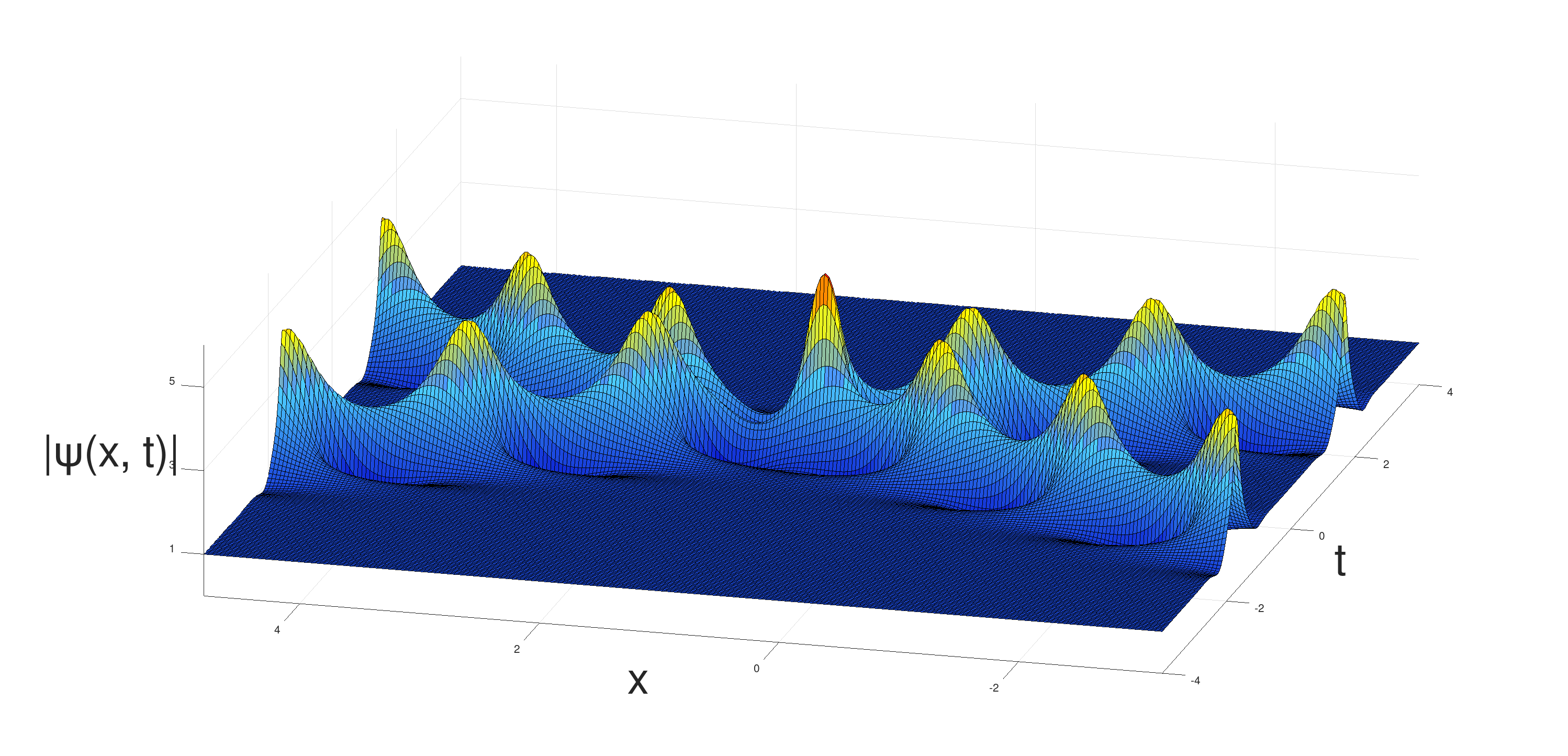}
					\caption{{ 2-fold domains on which the poles accumulate and the corresponding ``shielding" 2-breather. Top  left:
				 $\rho=\sqrt{6/5}$, $ d_0=\im\varsigma=3\im,\ d_1=-1$ and $r(z)=1$. Top right: $\rho=\sqrt{2},\ d_0=\im\varsigma=3\im,\ d_1=1$ and $m=2$.	Below the corresponding $2$-breather solutions with effective parameters $\lambda_{1,2}=(2\im, 4\im),\,,C_{1,2}=(-\frac35\im, \frac35\im)$ (left),  $\lambda_{1,2}=1\pm 3\im\, C_{1,2}=\pm1$	 (right) as per 			
				   \eqref{parameters}.)}}
				\label{2 fold 1}
\end{figure}

			We consider poles accumulating on $\mathcal{D}_1$, where now $\mathcal{D}_1$ is the $2$-fold  domain given by
				\begin{equation}
						\mathcal{D}_1=\left\{z \in \mathbb{C}\ s.t.\, \Big|\left(z-\im\varsigma\right)^2-d_1\Big|<\rho\right\},				
				\end{equation}
and $\mathcal{D}_2$ obtained by symmetry
\begin{equation}
						\mathcal{D}_2=\left\{z \in \mathbb{C}\ s.t.\, \left|\left(-\frac{1}{\overline{z}}-\im\varsigma\right)^2-d_1\right|<\rho\right\},
				\end{equation}
				with $\varsigma>1$ ($d_0=\im\varsigma$) and $d_1$ real.
			From the  equations \eqref{lambdak} and \eqref{cjshielding}, we obtain 
							\begin{equation}
							\label{parameters}
					\begin{split}
						&\lambda_{1}=i\varsigma+\sqrt{d_1},\;\;	\lambda_{2}=i\varsigma-\sqrt{d_1},\\
						&C_1=\dfrac{\rho^2r(\lambda_1)}{\lambda_1-\lambda_2},\quad 
						C_2=\dfrac{\rho^2r(\lambda_2)}{\lambda_2-\lambda_1}.				\end{split}
			\end{equation}
			Note that when $d_1<0$ the eingevalues $\lambda_1$ and $\lambda_2$ are purely imaginary so that one obtains   2-breather solution of Kutznesov-Ma type, while when $d_1>0$ one obtain a 
			   2-breather  solution of Tajiri-Watanabe  type because the eigenvalues $\lambda_1$ and $\lambda_2$  are complex.\\
  Two examples of breathers are obtained in Figure~\ref{2 fold 1} with  $r(z)=1$ .
%
			
					 \section{Conclusions and Outlook}
					 {
					 This paper was devoted to study 
					 a gas of $N$-breather solutions of the FNLS equation in a suitably taylored limit $N\to\infty$. Namely, we started from the well known Riemann-Hilbert problem leading to $N$-breather solutions \cite{Kuznetsov_1977}, \cite{Biondini_K_2014}  and considered families of such solutions with poles accumulating in specific domains, namely quadrature domains of a specific kind. Then we performed a specific $N\to\infty$ limit according to the following  prescriptions:
					 \begin{description}
\item[ i)] 
The soliton spectrum $\{\zeta_j\},\ j=1,\ldots, N$ is chosen to be uniformly distributed in  the ``parent'' domain  $\mathcal{D}_1$, the spectrum distribution in the other $three$ domains being obtained by the well-known symmetry properties of FNLS breather spectrum.
\item[ii)]
The norming constants $C_j$, $j=1,\dots N$  is interpolated by a smooth function $\beta_1(z,\bar{z})$ and  asymptotically scale with $1/N$.
 \end{description}
					   We had to carefully consider the so-called {\em theta condition}, and we restricted our study to the simplest case, namely the one for which 
$4\sum_{j=1}^N\arg(\zeta_j)=0$, which implies no phase shift for the $\pm\infty$ asymptotic values of the FNLS solution. 
This can be accomplished by cherrypicking a parent domain   $\mathcal{D}_1$   symmetric  with respect to  the positive imaginary axis. 

We showed that, under such circumstance, the {\em soliton shielding} phenomenon,  discovered in \cite{Bertola_G_O_2022} can happen also for gas of breathers.
Indeed, when the domain $\mathcal{D}_1$  is the $n$-fold quadrature domain given by the inequality $ \Big|(z-d_{0}  )^{n}-d_{1}\Big|<\rho\, , $ and the interpolating function  is of the kind $\beta_1(z,\bar{z})= n(\overline{z}-\overline{d_{0}})^{n-1}r(z)$ (for some
analytic function  $r(z)$), we have shown that the limiting solution is the $n$ Kuznetsov-Ma  and/or Tajiri-Watanabe  breathers. We illustrated our results in the simplest case with $n=1,2$.

It is fair to say that our results require some stringent conditions to hold. Unravelling the analytical properties of a general gas of breathers remains an open problem. 
In particular, the question on how to obtain general $n$ combinations of Kuznetsov-Ma and Tajiri-Watanabe shielding breathers by relaxing our condition of equal phases at $x=\pm\infty$ should be further examined.
Also, Akhmediev's breathers were not considered here.  For a gas of  such breathers, it is expected that the corresponding initial data be step-like oscillatory. Some preliminary steps along these lines have been undertaken in \cite{WZMZ25}.}

					\subsection*{Acknowledgments}
										This work has received funding by the Italian PRIN 2022 (2022TEB52W) - PE1 - project {\em The charm of integrability: from nonlinear waves to random matrices}. We also gratefully acknowledge the auspices of the GNFM Section of INdAM, under which part of this work was carried out, and the support of the project  MMNLP (Mathematical Methods in Non Linear Physics) of the Italian INFN.
C.P.'s research was supported by the Austrian Science Fund (FWF) [grant number Z 387-N].  We are also grateful to the anonymous referee for useful comments and remarks.
					\begin{appendices}
\section{Determination of the breather parameters}\label{AppA}
				The system of equations \eqref{sys_tot} is  further simplified, using the  symmetry of the matrix $M$
				    (see  formula 2.24 \cite{Biondini_K_2014})
				between the first column $M_1(z)$ and the second column $M_2(z)$:
									\begin{equation}
					\label{symmetry}
					M_1(z)=\dfrac{i}{z}M_2\left(-\frac{1}{z}\right).
					\end{equation}
				
				The symmetry condition \eqref{symmetry}  gives the following relation  	of  the matrix entries of $M(z)$ in \eqref{Martix M}
						\begin{equation}
			M_1(z)=\begin{bmatrix}
			1\\\frac{i}{z}\end{bmatrix}+\sum_{\alpha=1}^{2N}\dfrac{\begin{bmatrix}
			A_\alpha\\ B_\alpha\end{bmatrix}}{z-\zeta_\alpha}=\dfrac{i}{z}M_2\left(-\frac{1}{z}\right)=\begin{bmatrix}
			1\\\frac{i}{z}\end{bmatrix}-\sum_{\alpha=1}^{2N}\frac{i}{\overline{\zeta_\alpha}}\dfrac{\begin{bmatrix}
			E_\alpha\\ F_\alpha\end{bmatrix}}{z+\frac{1}{\overline{\zeta_\alpha}}},
			\end{equation}
			therefore
			\[
			\sum_{j=1}^{N}\dfrac{\begin{bmatrix}
			A_{j+N}\\ B_{j+N}\end{bmatrix}}{z-\zeta_{j+N}}=-\sum_{j=1}^{N}\frac{i}{\overline{\zeta_j}}\dfrac{\begin{bmatrix}
			E_j\\ F_j\end{bmatrix}}{z-\zeta_{j+N}}
			\]
		and
		\[
			\sum_{j=1}^{N}\dfrac{\begin{bmatrix}
			A_{j}\\ B_{j}\end{bmatrix}}{z-\zeta_{j}}=-\sum_{j=1}^{N}\frac{i}{\overline{\zeta_{j+N}}}\dfrac{\begin{bmatrix}
			E_{j+N}\\ F_{j+N}\end{bmatrix}}{z-\zeta_{j}},
			\]
or equivalently 
\begin{equation}
\begin{bmatrix}A_{j+N}\\ B_{j+N}\end{bmatrix}=-\frac{i}{\overline{\zeta_j}}\begin{bmatrix}
			E_j\\ F_j\end{bmatrix},\quad \begin{bmatrix}
			A_{j}\\ B_{j}\end{bmatrix}=-\frac{i}{\overline{\zeta_{j+N}}}\begin{bmatrix}
			E_{j+N}\\ F_{j+N}\end{bmatrix}.
			\end{equation}

		Inserting the above relation into the system \eqref{sys_tot} we obtain after some algebra the following linear system of  equations for the constants $E_\alpha$, $\alpha=1,\dots,2N$:
			
			\begin{equation}
			E_\alpha-\im \overline{c_\alpha}\sum_{j=1}^N\left(\frac{1}{\overline{\zeta_{j+N}}}\dfrac{E_{j+N}}{\overline{\zeta_\alpha}-\zeta_j}+
			\frac{1}{\overline{\zeta_{j}}}\dfrac{E_{j}}{\overline{\zeta_\alpha}-\zeta_{j+N}}\right)=-\overline{c_\alpha}.
			\end{equation}
						Using the  symmetries \eqref{symmetrie poli}, \eqref{biondini norming constants 1}, \eqref{biondini norming constants 2} and \eqref{simmetria theta}
 the above system of equations can be written in the form			
 \begin{equation}
 \label{A_N_breather}
			\begin{bmatrix}
			1_N+\overline{X}& -\overline{Y}D\\
			&\\
			D^{-1}Y&1_N+D^{-1}XD\end{bmatrix}\begin{bmatrix}E_1\\
			\vdots\\
			E_{2N}\end{bmatrix}=- \begin{bmatrix}\overline{c_1}\\
			\vdots\\
			\overline{c_{2N}}\end{bmatrix}
						\end{equation}
						where $1_N$ is the $N$-dimensional identity and 
						\[
						X_{kj}=\im\dfrac{c_k }{1+\zeta_k\zeta_j},\quad Y_{kj}=i\dfrac{c_k }{\zeta_k-\overline{\zeta_j}}, \;\;D=\mbox{diag}(\zeta_1,\dots,\zeta_N).
						\]
						Invertibility of the system of equations  \eqref{A_N_breather}
 follows from the solvability of the Riemann-Hilbert problem
						{B} 
			 for  the  $N$-breather   solution proved in Appendix \ref{AppB}.
						For $N=1$ the $1$-breather solution 							\eqref{Nbreathers}, 
				\begin{equation}
					\psi(x,t)=1-\im(E_1+E_2),
				\end{equation}
				is obtained by solving the  linear  system
				\begin{equation}
				\label{lin_2}
				\begin{bmatrix}
				1-\im \dfrac{\overline{c_1}}{1+\overline{\zeta_1}^2}&\im \dfrac{\overline{c_1}\zeta_1}{\overline{\zeta_1}-\zeta_1}\\
				&\\
				\im \dfrac{c_1}{\zeta_1(\zeta_1-\overline{\zeta_1})}&1+\im  \dfrac{c_1}{\zeta_1^2+1}
				\end{bmatrix}\begin{bmatrix}E_1\\
			&\\
			E_{2}\end{bmatrix}=- \begin{bmatrix}\overline{c_1}\\
			\\
			-\frac{c_1}{\zeta_1^2}\end{bmatrix}\,.
\end{equation}
	The solution   of such system for complex $\zeta_1\in D_1^+$ and  $c_1\in \C$  gives  the 	Tajiri-Watanabe  breather. In the case $\zeta_1$ is pure imaginary  and $ \zeta_1\in D_1^+$ one gets the Kutznesov-Ma breather.

			\section{Proof of Theorem~\ref{Theorem2.4}}\label{AppB}
			
			In this appendix we show the solvability of the Riemann-Hilbert problem {B} 
			 for $N$-breather   and the  
			 Riemann-Hilbert problem   {C} for the  breather gas.  We follow the steps in  \cite{Zhou}. Since the proof is identical in the two cases, we consider only the latter case. This provides a prove of Theorem \ref{Theorem2.4}.		The  Riemann-Hilbert problem  {C} can be written in the form 			
			\begin{equation}\label{eq:R-H_sol2 breather1}
			\left\{	\begin{split}
					&{Y}_+^\infty(z;x,t)-{Y}_-^\infty(z;x,t)=Y_-^\infty(z;x,t){J}^\infty(z;x,t),\quad z\in\boldsymbol{\gamma},\\
					& {Y}^\infty(z;x,t)= \mathbb{I}+\mathcal{O}\left(\frac{1}{z}\right),\quad\text{as}\ z\rightarrow \infty,\\
					& {Y}^\infty(z;x,t)= \frac{\im}{z} \begin{bmatrix}
						0 &1\\
						1&0
					\end{bmatrix}+\mathcal{O}\left(1\right),\quad\text{as}\ z\rightarrow 0,
				\end{split}\right.
			\end{equation}	
			which is equivalently written using the  Sokhotski–Plemelj integral formula  as
			\[
			{Y}^\infty(z;x,t)	= \mathbb{I}+ \frac{\im}{z}\sigma_2+\dfrac{1}{2\pi i}
			\int_{\boldsymbol{\gamma}} \dfrac{Y_-^\infty(s;x,t){J}^\infty(s;x,t)}{s-z}ds,
			\] 
			where we recall that $\boldsymbol{\gamma}$ is the union of all the contours $\gamma^{\pm}_1$ and $\gamma^{\pm}_2$.
			We can obtain an integral equation by taking the boundary value $ {Y}^\infty_{-}(z)$  as $z $ approaches  non tangentially  the  oriented contour $ \boldsymbol{\gamma}$  from the right:
\begin{eqnarray}
\label{eq:IntEq1} &&
 {Y}^\infty_{-}(\xi) =\mathbb{I}+ \frac{\im}{\xi}\sigma_2+ \lim_{\substack{z \to \xi \\ z \in \mbox{ \small right side of }  \boldsymbol{\gamma}}} \left( \frac{1}{2 \pi i} \int_{\boldsymbol{\gamma}} \frac{Y_-^\infty(s){J}^\infty(s)}{s-\xi} d s \right)  \ . 
\end{eqnarray}
We define  the integral operator $\mathcal C_{J^{\infty}}$ as
\begin{equation}
\label{eq:Cauchy_WN}
\mathcal C_{J^{\infty}}( h)(\xi)=\mathcal C_-( h J^{\infty})(\xi),
\end{equation}
where $\mathcal C_-$ is the Cauchy projection operator  acting on $L^2(\boldsymbol{\gamma},d s)\otimes Mat(2\times 2,\C)$  to itself, 
namely
\begin{equation}
\label{eq:Cauchy_op}
\mathcal C_-( h)(\xi) = 
\lim_{\substack{z \to \xi \\ z \in \mbox{ \small right side of }  \boldsymbol{\gamma}}}
\left( \frac{1}{2 \pi i } \int_{\boldsymbol{\gamma}} \frac{h(s)}{s-z} d s \right),
\end{equation}
here $L^2(\boldsymbol{\gamma}, d s)$ is the   the space of square integrable 
functions  on the contour $\boldsymbol{\gamma}$.
Then the integral equation \eqref{eq:IntEq1} takes the form 
\begin{eqnarray}
\label{E_-}
\left[  1 - \mathcal C_{J^{\infty}} \right]  {Y}^\infty_{-}(\xi) =  \mathbb{I}+ \frac{\im}{\xi}\sigma_2,
\end{eqnarray}
where $ 1$ is the identity in  $L^2(\boldsymbol{\gamma},\d s)\otimes Mat(2\times 2,\C)$.
Then Theorem 9.3 from \cite{Zhou}(p.984) guarantees that the operator 
$1 - \mathcal C_{J^{\infty}} $  is invertible as an operator acting from $L^2(\boldsymbol{\gamma}, d s)\otimes Mat(2\times 2,\C)$ to  itself. 
 Invertibility is guaranteed from the fact that $1 - \mathcal C_{J^{\infty}}$  is a Fredholm integral operator with zero index and the kernel
of $1  - \mathcal C_{J^{\infty}}$ is $\{0\}$. In particular this last point is obtained by applying the vanishing lemma \cite{Zhou}. Indeed suppose that exists $\hat{Y}_-\in L^2(\boldsymbol{\gamma},\d s)\otimes Mat(2\times 2,\C)$ such that
 $( 1  - \mathcal C_{J^{\infty}})\hat{Y}_-=0$. Then the quantity
 \begin{equation}
 	\hat{Y}(x,t;z)=\frac{1}{2\pi i}\int\limits_{\boldsymbol{\gamma}}\frac{\hat{Y}_-(s)J^\infty(s;x,t)}
{s-z}d s\end{equation}
solves the  following  Riemann-Hilbert problem:  
\begin{itemize}
\item $\hat{Y}(z;x,t)$  is analytic in $\C\backslash \boldsymbol{\gamma}$;
\item  $\hat{Y}_+(\xi)=\hat{Y}_-(\xi)(\mathbb{I}+J^\infty(\xi))$ for $\xi\in\boldsymbol{\gamma}$;
\item $\hat{Y}(z;x,t)={\mathcal O}(z^{-1})$ as $z\to\infty$. 
\end{itemize}
To show that such problem has only the zero solution we use the following properties:  \begin{itemize}
 \item  the jump matrix   satisfies the Schwartz symmetry: $J^\infty(z; x,t)=-\overline{ J^\infty(\ol z; x,t)}^t$;
  \item the contour $\boldsymbol{\gamma}$ is symmetric with respect to the real axis $\mathbb{R}$ (up to the orientation).
 \end{itemize}
 We define  $H(z)=\hat{Y}(z)\ol{\hat{Y}(\ol{z})}^t$, then clearly $\ol{H(\ol{z})}^t=H(z)$, $H(z)={\mathcal O}(z^{-2})$  as $z\to\infty$  and $H(z)$ is analytic in $\C\backslash\boldsymbol{\gamma}$.
We consider  the integral  on the real axis and apply contour deformation on the upper half space
\begin{align*}
\int_{-\infty}^\infty H_+(z)dz
&=\int_{\gamma^-_2\cup\gamma^+_1}H_-(z)dz\\
&=\int_{\gamma^-_2\cup\gamma^+_1}\hat{Y}_+(z)(\mathbb{I}+J^\infty(z))^{-1}((\mathbb{I}+\overline{J^\infty(\ol z)})^{-1})^t\ol{\hat{Y}_+(\ol{z})}^t dz\\
&=\int_{\gamma^-_2\cup\gamma^+_1}\hat{Y}_+(z)\ol{\hat{Y}_+(\ol{z})}^tdz=0\,.
\end{align*}
The second  identity is obtained using the boundary values of $\hat{Y}(z)$. In the  third identity  we have used the Schwartz reflection  of the jump matrix  $J^\infty$ and the fourth identity  is obtained using the fact that  $\hat{Y}(z)$ is analytic inside ${\gamma^-_2\cup\gamma^+_1}$.   We conclude that 
\[
0=\int_{-\infty}^\infty\hat{Y}_+(z)\ol{\hat{Y}_+(\ol{z})}^tdz=\int_{-\infty}^\infty\begin{bmatrix}
|\hat{Y}_{11}(z)|^2+|\hat{Y}_{12}(z)|^2&*\\
*&|\hat{Y}_{21}(z)|^2+|\hat{Y}_{22}(z)|^2
\end{bmatrix}dz,
\]
which implies $\hat{Y}_{ij}(z)=0$,  $i,j=1,2$  and  $z\in\mathbb{R}$. Due to analyticity of $\hat{Y}(z)$  in $\C\backslash\boldsymbol{\gamma}$ and the unique continuation theorem, then $\hat{Y}$ is identically zero. So the null space is empty and the operator   $1 - \mathcal C_{J^{\infty}}$ is invertible showing the existence of the  solution to the  Riemann-Hilbert problem for $Y^\infty(z;x,t)$.
The proof that the solution $\psi_\infty(x,t)$ is in $C^\infty(\R\times\R^+)$ relies on the proof of existence of the matrix 
 $\partial^n {Y}^\infty(z;x,t)$ for  $n \in \mathbb{N}$ with the following properties: 
\begin{enumerate}
	\item $ \partial^n  {Y}^\infty(z;x,t)$ is holomorphic for $z \in \C \setminus\boldsymbol{\gamma}$.
 	\item $ \partial^n  {Y}^\infty(z;x,t) = \mathcal O\left( z^{-1}\right)$,  as $z \to \infty$ 
	\item For $z \in \boldsymbol{\gamma}$, the boundary values of $ \partial^n  {Y}^\infty(z;x,t)$ satisfy the jump relation
	\begin{equation}
	\begin{gathered}
	\left(\partial^n  {Y}^\infty(z;x,t)\right)_+ = \left(\partial^n {Y}^\infty(z;x,t)\right)_- J^\infty(z,;x,t) + \mathcal{F}^{(n)}(z;x,t),  \\
	\mathcal{F}^{(n)}(z;x,t) := \sum_{\ell=1}^n \binom{n}{\ell} \partial^{n-j}  {Y}^\infty(z;x,t)  (z;x,t) \partial^{j} J^\infty (z;x,t).
	\end{gathered}
	\end{equation}
\end{enumerate}
Here  $\partial^n {Y}^\infty(z;x,t)$ is  the partial derivative with respect to a parameter, in our case $t$ or $x$. The details on the existence of the solution  to 
 such problem are given in  \cite{DZ},\cite{GGMN}.
\end{appendices}

				
\end{document}